\documentclass[10pt,conference]{IEEEtran}
\usepackage{cite}
\usepackage{amsmath,amssymb,amsfonts}
\usepackage{algorithmic}
\usepackage{graphicx}
\usepackage{textcomp}
\usepackage{xcolor}
\usepackage[hyphens]{url}
\usepackage{fancyhdr}
\usepackage{hyperref}
\usepackage{booktabs}
\usepackage{tikz}
\usepackage{pifont} %
\usepackage{subcaption}
\usepackage[inline]{enumitem}
\usepackage{xcolor}

\newcommand{\hpcayear}{2025}

\title{Revisiting Reliability in Large-Scale Machine Learning Research Clusters}

\def\hpcacameraready{} %

\newcommand\hpcaauthors{
Apostolos Kokolis*, Michael Kuchnik*, John Hoffman, Adithya Kumar,\\ Parth Malani, Faye Ma, Zachary DeVito,  Shubho Sengupta, Kalyan Saladi, Carole-Jean Wu
}
\newcommand\hpcaaffiliation{FAIR at Meta}
\newcommand\hpcaemail{\{akokolis, mkuchnik, 
johnhoffman, 
kadithya,
pmalani,
fms,
zdevito,
ssengupta,
skalyan,
carolejeanwu\}@meta.com}

\definecolor{lightred}{rgb}{1, 0.0, 0.1}
\definecolor{darkgreen}{rgb}{0.067, 0.8, 0.4}

\newcommand\ettrname{Effective Training Time Ratio}
\newcommand\ettrabbr{ETTR}

\newcommand{\xmark}{\ding{55}}
\newcommand{\rxmark}{\color{lightred}\xmark}
\newcommand{\gcheckmark}{\color{darkgreen}\ding{51}}
\newcommand*\circled[1]{\tikz[baseline=(char.base)]{%
            \node[shape=circle,draw,inner sep=1pt] (char) {#1};}}

\newcounter{lessoncounter}
\newcommand\lesson[1]{\noindent\stepcounter{lessoncounter}\textbf{Observation \arabic{lessoncounter}:} #1}

\newboolean{expandedversion}
\setboolean{expandedversion}{true} %

\newboolean{isarxivversion}
\setboolean{isarxivversion}{true} %

\newboolean{fontcompat}
\setboolean{fontcompat}{false} %

\newcommand{\arxivonly}[1]{%
  \ifthenelse{\boolean{expandedversion}}{{%
  \color{black}%
     #1%
  }}{}%
}

\newcommand{\hpcaonly}[1]{%
  \ifthenelse{\boolean{expandedversion}}{}{{
     #1
  }}
}

\author{
  \ifdefined\hpcacameraready
    \IEEEauthorblockN{\hpcaauthors{}}
      \IEEEauthorblockA{
        \hpcaaffiliation{} \\
        \hpcaemail{}
      }
  \else
    \IEEEauthorblockA{Apostolos Kokolis*, Michael Kuchnik*, John Hoffman, Adithya Kumar,\\ Parth Malani, Faye Ma, Zachary DeVito,  Shubho Sengupta, Kalyan Saladi, Carole-Jean Wu\\ \\
    FAIR at Meta}
    
  \fi 
}

\fancypagestyle{camerareadyfirstpage}{%
  \fancyhead{}
  
  \fancyhead[C]{
    \ifdefined\aeopen
    \parbox[][12mm][t]{13.5cm}{\hpcayear{} IEEE International Symposium on High-Performance Computer Architecture (HPCA)}    
    \else
      \ifdefined\aereviewed
      \parbox[][12mm][t]{13.5cm}{\hpcayear{} IEEE International Symposium on High-Performance Computer Architecture (HPCA)}
      \else
      \ifdefined\aereproduced
      \parbox[][12mm][t]{13.5cm}{\hpcayear{} IEEE International Symposium on High-Performance Computer Architecture (HPCA)}
      \else
      \parbox[][0mm][t]{13.5cm}{\hpcayear{} IEEE International Symposium on High-Performance Computer Architecture (HPCA)}
    \fi 
    \fi 
    \fi 
    \ifdefined\aeopen 
      \includegraphics[width=12mm,height=12mm]{ae-badges/open-research-objects.pdf}
    \fi 
    \ifdefined\aereviewed
      \includegraphics[width=12mm,height=12mm]{ae-badges/research-objects-reviewed.pdf}
    \fi 
    \ifdefined\aereproduced
      \includegraphics[width=12mm,height=12mm]{ae-badges/results-reproduced.pdf}
    \fi
  }
  \fancyfoot[C]{}
}
\begin{document}
\maketitle

\ifdefined\hpcacameraready 
  \thispagestyle{camerareadyfirstpage}
  \pagestyle{empty}
\else
  \thispagestyle{plain}
  \pagestyle{plain}
\fi

\newcommand{\hpcaheight}{0mm}
\ifdefined\eaopen
\renewcommand{\hpcaheight}{12mm}
\fi 

\begin{abstract}
Reliability is a fundamental challenge in operating large-scale machine learning (ML) infrastructures, particularly as the scale of ML models and training clusters continues to grow.
Despite decades of research on infrastructure failures, the impact of job failures across different scales remains unclear.
This paper presents a view of managing two large, multi-tenant ML clusters, providing quantitative analysis, operational experience, and our own perspective in understanding and addressing reliability concerns at scale.
Our analysis reveals that while large jobs are most vulnerable to failures, smaller jobs make up the majority of jobs in the clusters and should be incorporated into optimization objectives. We identify key workload properties, compare them across clusters, and demonstrate essential reliability requirements for pushing the boundaries of ML training at scale.

We hereby introduce a taxonomy of failures and key reliability metrics, analyze 11 months of data from two state-of-the-art ML environments with 4 million jobs and over 150 million A100 GPU hours.
Building on our data, we fit a failure model to project Mean Time to Failure for various GPU scales. We further propose a method to estimate a related metric, \ettrname{}, as a function of job parameters, and we use this model to gauge the efficacy of potential software mitigations at scale.
Our work provides valuable insights and future research directions for improving the reliability of AI supercomputer clusters, emphasizing the need for flexible, workload-agnostic, and reliability-aware infrastructure, system software, and algorithms. 

 \end{abstract}

\section{Introduction}
\renewcommand*{\thefootnote}{}
\footnotetext{*Equal Contribution.}
\renewcommand*{\thefootnote}{\arabic{footnote}}
Accelerating innovations towards Artificial General Intelligence is pushing the capability of today’s computing infrastructure, demanding system breakthroughs for model training at-scale.
Companies are investing in large-scale clusters with thousands of GPUs and fast interconnect networks.
For example, Meta introduced its AI supercomputer, interconnecting 2,000 NVIDIA DGX A100 systems (16,000 A100 GPUs)~\cite{ai_rsc} with a 1600 Gb/s InfiniBand network and petabytes of storage capacity~\cite{ai_rsc}. Within just two years, Meta debuted two 24,000-GPU training clusters to accelerate generative AI technology development~\cite{meta_genai_cluster}. 
At the same time, Google invested in the next generation of Tensor Processing Units (TPUs) that form the foundation of its AI supercomputers~\cite{tpuv1,tpuv4}. Each TPU v5p pod constitutes 8,960 chips, which communicate via the inter-chip interconnect of 4,800 Gb/s/chip in a 3D torus topology~\cite{tpuv5}.
Such heavy infrastructure investments inevitably stress the limits of the existing systems stack.

With the rise of Large Language Models (LLMs)---Megascale~\cite{megascale}, LLaMa~\cite{touvron2023llamaopenefficientfoundation}, Gemini~\cite{gemini}, GPT4~\cite{gpt4}---ML training shifted the scale of a single training job from tens to tens of thousands of accelerators, presenting concrete use-cases for the aforementioned investments.
At such scale, failures are not \textit{a matter of if, but a matter of when}.
Thus, system design entails new and unexpected challenges in the operation, efficiency, and reliability of a cluster, creating the need for new solutions.
Indeed, hardware failures~\cite{gemini,llama3-model-card} are just some of the issues that are individually rare yet become increasingly likely at scale, involving solutions that span from hardware to the design of a machine learning cluster scheduler.

In this paper, we present our infrastructure experience toward training a plethora of large-scale models, including earlier foundation models~\cite{touvron2023llamaopenefficientfoundation} as their usage became prevalent, with the largest jobs utilizing 4k GPUs or more.
Unlike prior work, our hardware and software infrastructure is tailor-designed to be capable of serving a diverse set of workloads---4k GPU jobs constitute less than 1\% of our jobs while consuming 12\% of the GPU resources at the cluster level.
Our experience catering to both large- and small-scale jobs demonstrates diversity in infrastructure needs that is rarely observed in more specialized clusters devoted to only LLMs.

Understanding the underlying causes of job failures---let it be hardware, system software, applications, or some combinations of the above---is key to improving training reliability and advancing large model development. 
In this paper, we present 11 months of data collected from state-of-the-art AI research clusters with \textgreater80\% utilization. The results based on real-world training systems highlight the diversity of research workloads across 2 clusters, spanning 24k of NVIDIA A100 GPUs. 
Our primary focus is on job-level failures.
We primarily view failures through the lens of the \textit{scheduler} and \textit{server-level health checks}.
We additionally provide some network-level reliability experience.
Finally, we share lessons we have learned in mitigating failures at scale, tracking reliability metrics, making infrastructure changes, and diagnosing common application pitfalls, culminating in suggestions for future opportunities.
\arxivonly{In doing so, we provide and analyze server-level component failure rates, including Mean Time to Failure (MTTF) projections.}

To the best of our knowledge, we present the first infrastructure analysis of ML research workloads at the $10^5$ GPUs scale.
Our contributions include the following:
\begin{enumerate}[leftmargin=*]
    \item \textbf{Introducing a failure taxonomy and key reliability metrics} that we use in operating the clusters, which cater to minimal incidental complexity and maximum flexibility in running ML workloads ranging from 1 to 4k+ GPUs.
    \item \textbf{Pinpointing reliability improvement opportunities based on an analysis of deployed ML training systems.} The 11-month data from training various ML jobs in two state-of-the-art machine learning environments spans 4 million jobs and over 150 million A100 GPU hours. We find that jobs in our research cluster are more diverse than implied by LLM workloads, motivating workload agnostic infrastructure techniques.
    \item \textbf{Validating projections of Mean Time to Failure for various GPU scales based on failure data.} Our predictions are in agreement with theory and are validated on job data up to 4k GPUs in scale.
    \item \textbf{Designing and validating an analytical estimator for expected \ettrname{}} as a function of various job parameters using several aggregate statistics from our data. 
    This approach is general across other clusters and workloads.
    \item \textbf{Proposing and evaluating software mitigations for infrastructure issues affecting AI supercomputing clusters}, including experience with adaptive routing, health checks, and faulty node detections.
    We use our experience to project how failures may impact future workloads.
\end{enumerate}

In the rest of this paper, we provide an overview of our cluster in \S\ref{sec:system_infra}, we dive into failure data in \S\ref{sec:analysis}, we propose mitigations in \S\ref{sec:cluster-optimization}, and we close with future directions in \S\ref{sec:opportunities}) and related work in \S\ref{sec:related_work}. %
\section{System Infrastructure}%
\label{sec:system_infra}

In this section, we describe how workloads influence the design of our clusters.
While clusters can be specialized to optimize toward a specific workload, research clusters are, by definition, expected to have constantly changing workloads with potentially unforeseen needs.
Therefore, we believe that research clusters should be general, maximize productivity, and minimize incidental complexity.
Our two sister clusters, \texttt{RSC-1}{} and \texttt{RSC-2}{}, follow the same design template discussed below.
\texttt{RSC-1}{} is a general ML cluster (e.g., training some of the prominent LLMs) of 16k GPU size, while \texttt{RSC-2}{} focuses on vision applications and is of 8k GPU size.
As we discuss later (\S\ref{sec:analysis}), the workload differences manifest in different usages---for example, workloads on  \texttt{RSC-2}{} have a significant tilt towards 1-GPU jobs, along with jobs going up to 1k GPU size.

\subsection{Scheduler and Storage Infrastructure Overview} 
\label{sec:infrastructure}
\begin{figure}[t]
\hbox{\hspace{12pt}
\includegraphics[width=0.8\columnwidth]{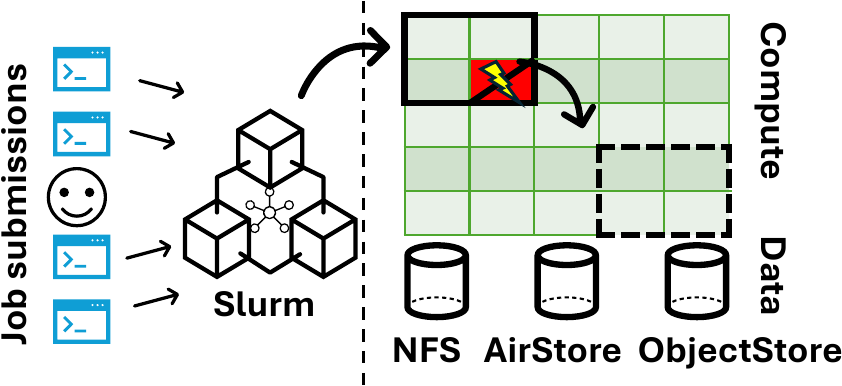}}
\caption{System Overview of the Research Cluster.}
\label{fig:rsc_frontend_backend}
\end{figure}

Our design of \texttt{RSC-1}{} and \texttt{RSC-2}{} prioritized ease of use while favoring simplicity.
The benefit of our design is that the entire stack is mature and does not require extensive custom datacenter designs, reducing our time-to-market to only 1.5 years.
Additionally, we aim to provide users with the requested number of GPUs with no strings attached to maximize productivity---users do not have to deal with complexity in the form of novel hardware or virtualization. Figure \ref{fig:rsc_frontend_backend} provides an overview of how users interact with both our clusters.
Users submit a \textit{job}, which is comprised of many \textit{tasks}, each of which can run on the GPUs of a \textit{node}.

\textbf{Scheduler.} Leaning into the High-Performance Computing (HPC) stack, our clusters use the Slurm~\cite{slurm} scheduler on top of bare-metal allocations.
The cluster is configured such that groups of users have a maximum quota of GPUs that is determined by a project-specific allocation.
Users submit jobs using shell scripts (\texttt{sbatch}) or Python wrappers (\texttt{submitit}~\cite{submitit}).
Slurm, in turn, attempts to co-locate the tasks given the physical network topology.
Jobs are eligible to be preempted after running for two hours, and they have a maximum lifetime of seven days.
Slurm attempts to schedule jobs based on priority order, which is a function of many variables, including the project's allocation and the job's age~\cite{SLURMPriority}.

ML workloads follow \textit{gang scheduling} semantics. 
Gang scheduling ensures that all required resources are allocated simultaneously across multiple tasks.
This coordination is essential for optimizing performance and efficiency in large-scale ML workloads.
However, as shown in Figure~\ref{fig:rsc_frontend_backend}, a single task failure can force a complete re-allocation of the job.
This motivates \textit{fault tolerance} strategies, such as checkpointing and redundancy, to be used for gang scheduling.
Checkpointing allows a job to recover from a saved job state, minimizing the impact on overall job progress, while redundancy reduces the likelihood of a job failure, minimizing the rate of failures.

Users who submit a job are given a guarantee by our infrastructure---if a failed health check results in a terminated job, the system automatically requeues the job, with the same Job ID, as shown in Figure~\ref{fig:rsc_frontend_backend}.
Overall, our clusters average 7.2k for \texttt{RSC-1}{} and 4.4k for \texttt{RSC-2}{} jobs submitted per day, averaging 83\% and 85\% cluster utilization, respectively.

\textbf{Storage.} Input and output data, as well as checkpoints of a job are expected to be durable and decoupled from the lifetime of the particular job.
Our clusters have three storage offerings:
\begin{enumerate*}[label=\protect\circled{\arabic*}]
\item a POSIX-compliant storage offering backed by flash storage and exported through the NFS protocol,
\item a custom, high bandwidth dataset-focused offering, AirStore, 
and
\item an object-storage of high capacity and throughput, ObjectStore.
\end{enumerate*}
The first facilitates ease of use, providing users with home directories, Python environments, and the ability to perform read and write operations for common patterns such as checkpointing. 
For the second, dataset access is accelerated using a custom high-performance read-only caching service, AirStore, also backed by bulk flash storage. 
Finally, we have an object-storage interface (ObjectStore) for checkpointing and storing files when the NFS endpoint is insufficient. 
Checkpointing is of paramount importance in the interest of fault tolerance.
The availability of multiple options enables users to interpolate between ease of use and performance.

\begin{figure}
\includegraphics[width=\linewidth]{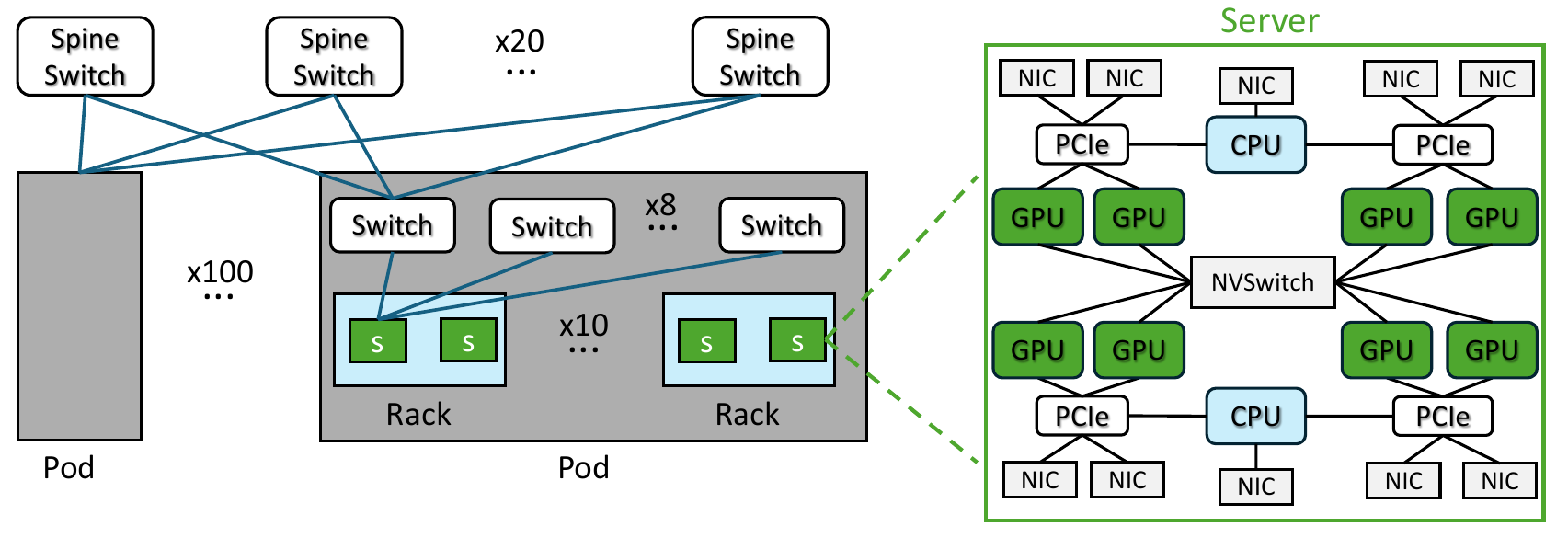}
\caption{The Network Topology of \texttt{RSC-1}{} (similar for \texttt{RSC-2}{}).}
\label{fig:rsc_architecture}
\end{figure}

\subsection{Compute and Network Infrastructure Overview} 

An HPC cluster's core hardware components are \textit{compute}, \textit{networking}, and \textit{storage} (discussed above).
Users provide the instructions to utilize these components via the \textit{jobs} that they submit to the \textit{scheduler}. The topology of our clusters is shown in Figure \ref{fig:rsc_architecture}, where the system layout of the nodes as well as the contents of a single server are presented.

\textbf{Compute.} Both the clusters we present in this paper are bare-metal, DGX~\cite{DGXA100Datasheet} based clusters with Dual AMD Rome 7742 CPUs and $8\times$ NVIDIA A100 80GB per server (node).
The GPUs are connected via a high-bandwidth NVSwitch.

\textbf{Networking.} In practice, hundreds of servers can be used in a job. The servers are connected with two types of interconnects, \textit{front-end} and \textit{back-end}.
The front-end network manages control-plane (i.e. scheduling and TCP connections) and storage traffic via Ethernet. 
Meanwhile, the back-end network uses an Infiniband fabric for low-latency model gradient exchange during neural network training.
Servers are connected via a rail-optimized Infiniband backend network, some of which is shown in Figure~\ref{fig:rsc_architecture}.
The rail-optimized topology means that GPUs of the same local server rank are locally connected, bypassing one level of switches.
Communication is grouped into logical domains: each rack has two servers, and ten racks are connected via a rail-optimized network, forming a \textit{pod}.
Pod-pod communications going through the next level of switches (spine switches).

The scheduler and the model training framework (e.g., PyTorch~\cite{pytorch2}) are expected to abstract out much of the complexity of the networks---offering a traditional collective-based communication model that should be portable and efficient across a variety of potential job allocations.
Crucially, the backend network software is able to exploit locality if it exists (e.g., opting to use high-bandwidth NVSwitch over Rail-connected Infiniband links over top-of-rack switches).
As we discuss below (\S\ref{sec:opportunities}), today's HPC-style collectives do, however, come with drawbacks.

\subsection{Observations on Cluster Infrastructure}
\lesson{\textit{Cluster uptime is critical.} Our clusters are fully loaded. Any downtime results in excessive queueing and is considered a major event. The cluster must adapt to failures online and ideally auto-requeue infrastructure-related failures.}

\textbf{Health Checks.} \label{sec:health-checks}
Because of the gang scheduling semantics of ML jobs, failures have a large effect on the reliability of an entire job---a single failure of a system component can cause thousands of GPUs to sit idle. 
Importantly, at the scale at which our clusters are operating, the time between component failures may be small enough to be disruptive.
Because of the large scope of potential failures and the overhead associated with transparently recovering from them,
our infrastructure is instead designed to check that jobs are running on healthy hardware, restarting the job on different nodes if there is a failure.
This can be viewed as a cooperative recovery strategy as the application is still responsible for correctly implementing checkpoint and resume logic.

To find, detect, and remove failed nodes, the scheduler responds to a series of \textit{health checks} run periodically on each node in the cluster.
\arxivonly{We analyze job failures through these health checks in \S\ref{sec:analysis}.}
A core philosophy underlying our training cluster design is to strive for \textit{\textbf{no second job failure from a bad node}}---a failed node is a poor scheduling candidate.

Slurm can run checks before and after a job runs~\cite{SLURMPrologEpilog}.
Moreover, we have health checks that are periodically scheduled to run every five minutes and return codes indicating success, failure, or warning.
Each health check examines some aspect of node health, spanning from GPU errors (e.g. XID errors~\cite{xid}) to file system mounts, and services status (i.e., scheduler).
Note that checks can have overlapping signals into a failure domain.
For example, a PCIe failure indicates that the GPU is inaccessible, even if the GPU did not incur the corresponding XID event itself. This situation occurs in our logs 57\% of the time on \texttt{RSC-1}{} (37\% on \texttt{RSC-2}{}).
Therefore, even if one check does not fire when it should, another overlapping check would hopefully catch the failure.
The most extreme case of this is \texttt{NODE\_FAIL}, which acts as a catch-all via Slurm heartbeats when a node becomes unresponsive to other health checks that are running on the node itself.

Periodic health checks are essential to prevent repeated job failures from the same unhealthy nodes.
The checks are tuned to have a low false positive rate---they have been previously calibrated such that less than 1\% of successfully completed jobs observe a failed health check, though note that we can only observe correlations and not causations.  %

Different checks have different severity. 
High severity check failures will immediately signal a scheduler handler to remove the node and reschedule all jobs executing on the node, while lower severity checks will signal to the scheduler to remove the node for remediation after jobs running on the node have finished, successfully or unsuccessfuly.
In the first category of check are the following: GPU not accessible, an NVLink error, uncorrectable ECC, failed row-remaps, PCI or IB link errors, block device errors, and missing mountpoints.
Nodes that are not failing any health checks are available for scheduling jobs. 
When a health check fails for a node, the node will transition to a remediation state and will become unavailable for scheduling until it is fixed and all checks are passing.
The transition to remediation can happen either immediately (for high severity checks) or after the current job finishes.

Health check importance can be motivated with the counterfactual of scheduling on possibly unhealthy nodes. Even if a small fraction of nodes are unhealthy, the probability of a job occupying an unhealthy nodes increases exponentially at scale.
We emphasize that health checks are the first-line defense toward ensuring a reliable computational substrate---though applications must continue to be proactive---which brings us to the next lesson.

\lesson{\textit{One bad node spoils the bunch.} Health checks prevent correlated failures due to repeated scheduling on defective nodes (``restart loops''). The inability to remove such nodes from capacity would result in the inability to effectively run large, gang-scheduled jobs and would severely cripple the cluster's efficiency. Recovering from random failures is only effective once defective nodes can be reliably rotated out.}

\subsection{Metrics}
\label{sec:metrics}
There are three critical metrics we consider in this paper for understanding how an ML cluster is performing: \ettrname{} (\ettrabbr{}), %
Goodput, and Mean Time to Failure (MTTF).

\textbf{\ettrname{} (\ettrabbr{}).} 
\ettrabbr{} is defined as the ratio of \emph{productive runtime} to \emph{available wallclock time} of a \emph{job run{}}. A \emph{job run{}} consists of one or more scheduler jobs related to the same \textit{logical job}~\cite{google_trace_2012}. For example, a multi-week LLM pretraining run may consist of multiple different jobs demarcated by pre-emptions and infrastructure failures (\ettrabbr{} attempts to ignore the impact from userspace failures to focus on impact from cluster stability only). The \emph{available wallclock time} of a job run{} is defined as the total time a job in the multi-job run{} was either
\begin{enumerate*}[label=\protect\circled{\arabic*}]
\item scheduled or
\item eligible to be scheduled but waiting in the queue.
\end{enumerate*}
\emph{Productive runtime} refers to scheduled time during which meaningful progress is being made for the workload.
The exact definition of \textit{productive runtime} is open to interpretation depending on context, but we consider three sources of unproductive scheduled time: 

\begin{enumerate}[leftmargin=*] %
    \item \textbf{Catching up from last saved checkpoint}: Re-training between the most recent checkpoint and a job interruption.
    \item \textbf{Restart overhead}: all initialization tasks that need to be performed after a restart that wouldn't otherwise be needed.
    \item \textbf{Checkpoint overhead}: The time checkpointing adds to job runtime.
\end{enumerate}
\arxivonly{All of these are highly job dependent, and we currently lack a reliable way for tracking either at scale with confidence. 
However, we treat these as free parameters to explore, filling in with reasonable values we have encountered anecdotally in collaborating with various research teams.}
\hpcaonly{All are job dependent. We treat them as free parameters and empirically fill them with values we measured in our clusters.}

\ettrabbr{} varies from 0 (the job never makes any meaningful progress) to 1 (100\% of the wallclock time was spent making meaningful progress i.e., no queueing or unproductive runtime).
\ettrabbr{} is similar to the canonical \textit{job slowdown} metric \cite{Harchol-Balter-2002}, defined as the ratio between wallclock time and the amount of scheduled time for a given job. However, \ettrabbr{} additionally accounts for unproductive runtime and inverts the ratio for arguably better interpretability.

\ettrabbr{}-like metrics, such as tracking job runtime until failure, were initially used for tracking the training efficiency of our LLMs\arxivonly{, like LLaMa~\cite{touvron2023llamaopenefficientfoundation},} as infrastructure issues were iteratively diagnosed.
Since then, such metrics were generalized to \ettrabbr{} and continue to be useful for more recent LLMs~\cite{llama3-model-card} outside of the presented clusters. For instance, Google Cloud defines a similar metric that they term ``Runtime Goodput''~\cite{ml_productivity_goodput}. To differentiate between the goodput metric we refer to in this paper (which only includes impact from wasted compute and not from e.g. wait time), we use the term \textbf{\ettrname{} (\ettrabbr{})}.
Similarly, note that the \ettrabbr{} we use differs from other definitions~\cite{Gemini_Wang2023} in that we model the wait time found in multi-tenant clusters.

Other potential metrics for characterizing model performance include Model Flops Utilization (MFU)~\cite{korthikanti2022reducingactivationrecomputationlarge,palm}, which we leave out of this paper.
MFU corresponds to the number of FLOPs a model theoretically utilizes compared to the hardware peak FLOPs, making it difficult to apply generally across an entire cluster.

\textbf{Goodput.} ETTR and MFU can be viewed as per-job efficiency metrics. The cluster as a whole can be measured in terms of \textit{goodput}, which is the amount of productive work completed in aggregate per unit time.
The goodput can be normalized by the maximum possible goodput to produce a utilization in the range 0 to 1.
The clusters discussed in this paper operate at high utilization (so potential goodput is %
limited more by capacity rather than available work), and thus job preemption, resource fragmentation, and failures are the dominant sources of lost goodput. While we use goodput to communicate loss in certain restricted scenarios in this paper, we focus on \ettrabbr{} as the main measure of job productivity on our clusters.

\textbf{Mean Time to Failure (MTTF).}
A key statistic in any reliability study is the Mean Time to Failure  (MTTF), a measure of how often failures occur.
It is the amount of measured system time divided by the amount of failures.
The \textit{failure rate} is the inverse.
The MTTF ranges from 0 to $\infty$ and gets smaller as sources of failure sum to higher total failure rates and thus lower MTTF.
MTTF can be used to configure the optimal checkpoint strategy under nonzero checkpoint overhead~\cite{Daly_2006, Young_1974}.

\begin{table*}[h!]
\centering
\begin{tabular}{l c c c l}
\toprule
Failure Symptoms & \multicolumn{3}{c}{Failure Domain} &  Likely Failure Cause \\
& User Program & System Software & Hardware Infra &  \\
\midrule
OOM & \gcheckmark & \rxmark & \rxmark & User Bug \\
GPU Unavailable & \rxmark & \gcheckmark & \gcheckmark & PCIe error, Driver/BIOS, thermals \\
GPU Memory Errors & \rxmark & \rxmark & \gcheckmark & Thermal Noise, Cosmic Rays, HBM Defect or Wear \\
GPU Driver/Firmware Error & \rxmark & \gcheckmark & \rxmark & Outdated Software, High Load \\
GPU NVLink Error & \rxmark & \rxmark & \gcheckmark & Electro/Material Failure, Switch \\
Infiniband Link & \rxmark & \rxmark & \gcheckmark & Electro/Material Failure, Switch \\
Filesystem Mounts  & \rxmark & \gcheckmark & \rxmark & Failed Frontend Network, Drivers in D State, Storage Backend \\
Main Memory Errors & \rxmark & \rxmark & \gcheckmark & Circuit Wear, Thermal Noise, Cosmic Rays  \\
Ethlink Errors & \rxmark & \rxmark & \gcheckmark & Electro/Material Failure, Switch \\
PCIe Errors & \rxmark & \rxmark & \gcheckmark & GPU Failure, Poor Electrical Contacts \\
NCCL Timeout & \gcheckmark & \gcheckmark & \gcheckmark & Userspace Crash, Deadlock, Failed HW  \\
System Services & \gcheckmark & \gcheckmark & \gcheckmark & Userspace Interference, Software Bugs, Network Partition  \\
\bottomrule
\end{tabular}
\vspace{1em}
\caption{Taxonomy of Failures. Users and cluster operators must infer a cause from a potentially ambiguous symptom. A common error is to misattribute the cause to the wrong component, especially when multiple domains are suspect. %
}
\label{tab:failure_taxonomy}
\end{table*}

\subsection{Failure Taxonomy}
\label{sec:failure-taxonomy}
Failure attribution is the process of assigning blame for a job failure to a cause. Our experience indicates that failure attribution is a challenging and noisy process. For instance, NCCL timeouts are a relatively common occurrence~\cite{he2023unicroneconomizingselfhealingllm}. In PyTorch, a NCCL timeout occurs whenever a rank observes that a collective operation, such as an \texttt{All-Reduce}, has not completed within several minutes.  While this can mean that some network issue occurred, it can also mean that some other rank simply never started that same operation because it was, for example, stuck trying to load data for the next iteration. In this case, the rank that times out is fully functional. 
The rank at fault may itself be unresponsive either due to a user software or due to infrastructure error (which itself can occur at a link or switch level). Tracing the root cause from user-level stack traces would require potentially many layers of precise and distributed logging, spanning from the ML application down to distributed collectives and low-level infrastructure.

Thus, our failure taxonomy, shown in Table~\ref{tab:failure_taxonomy}, is based on the principle that there may be many potential root causes for any given symptom, and the only way to limit the hypothesis space is to rule out unlikely causes.
We therefore propose to diagnose and root cause errors by \textit{differential diagnosis} over \textit{failure domains}---using a variety of performance indicators to flag where errors could have occurred, thus limiting a specific failure to a small subset of possible causes.

Our failure domains cover user code, system software (e.g., drivers, PyTorch, OS), and hardware (the components presented in \S\ref{sec:system_infra}).
Similar to prior work~\cite{jeon2019}, we observe that symptoms can map to multiple failure domains.
In a typical case, users should ensure their program does not have an obvious bug.
From a cluster operator point of view, hardware errors must be further binned by being transient (e.g., ECC error, link flap) or permanent (e.g., degraded hardware that requires repair or replacement by a vendor).
The tracking of information relevant to this failure taxonomy must be managed automatically (e.g., by health checks \S\ref{sec:infrastructure}), since 1) the pairing of program to machines is nondeterministic and 2) failures are often rare events.

We find that having an abundance of information covering various aspects of hardware and system software allows us to more quickly determine what caused a particular set of symptoms.
In some cases, it may even be expected that multiple concurrently firing health checks point to the same error (e.g., PCIe events may affect the GPU).

\lesson{\textit{Beware of the red-herrings.} Errors with multiple potential causes are difficult to diagnose. Errors such as NCCL timeouts may be naively attributed to a proximal cause e.g., on the network rather than a deadlock. Networking has a large ``blast-radius'', causing errors across the stack.
Some errors are transient and will fail to consistently reproduce---manifesting as statistical processes that can be observed via fleet-wide health checks.
Other errors are correlated to specific node hardware and may become more likely as they occur.
Table~\ref{tab:failure_taxonomy} summarizes our taxonomy and experience.
}

\section{Understanding Lay of the Land of Large-Scale ML Training Clusters}
\label{sec:analysis}

Our analysis is based on two research clusters and spans 11 months of measurement data.
It builds on the terminology of the Slurm scheduler and the node-level health-checks discussed previously (\S\ref{sec:infrastructure}).
Note that the clusters discussed in this section are over-provisioned, and project-level QoS and allocations are major factors in determining which jobs run.

\begin{figure}
  \centering
  \ifthenelse{\boolean{fontcompat}}{{%
  \includegraphics[width=1.\columnwidth]{figures/job_status_rsc_percf.pdf}%
  }}{{%
  \includegraphics[width=1.\columnwidth]{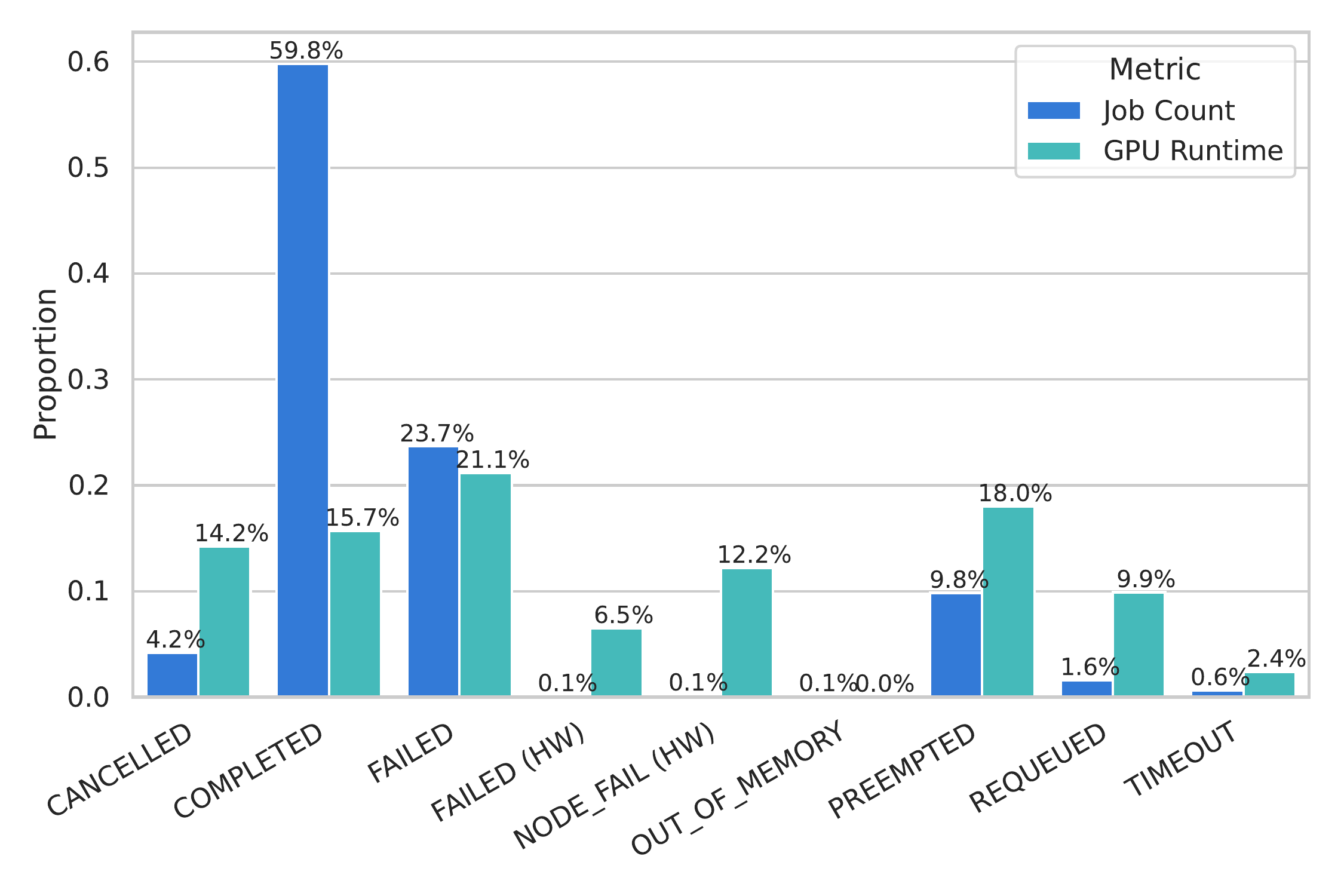}%
  }}
  \caption{Scheduler Job Status Breakdown by Number of Jobs and GPU Runtime on \texttt{RSC-1}{}.}
  \label{fig:scheduler-status-breakdown}
\end{figure}

\vspace{0.25cm}
\noindent{\textbf{Scheduler Job Status Breakdown.} A Slurm job can be \texttt{CANCELLED}, \texttt{COMPLETED}, \texttt{OUT\_OF\_MEMORY},  \texttt{FAILED} because the application returned a non-zero exit code, \texttt{NODE\_FAIL} because of a faulty node, \texttt{PREEMPTED} in favor of a higher priority job, \texttt{REQUEUED}, or \texttt{TIMEOUT}.
Figure~\ref{fig:scheduler-status-breakdown} illustrates the scheduler job status breakdown for the \texttt{RSC-1}{} cluster.
60\% of scheduled jobs completed.
24\% and 0.1\% of jobs failed because of \texttt{FAILED} and \texttt{NODE\_FAIL}, respectively.
10\% of jobs were pre-empted, 2\% requeued, 0.1\% ran out of memory, and 0.6\% timed-out}. 

Looking at infrastructure related failures, marked with \texttt{(HW)} in Figure~\ref{fig:scheduler-status-breakdown}, we see that such failures affect 0.2\% of jobs.
Nevertheless, we see that 18.7\% of runtime is impacted by these failures.
As we shall discuss below (Figure~\ref{fig:job-size-distribution-submitted}), this is not surprising given that we expect infrastructure failures to impact large jobs, which are rare by absolute number of jobs but occupy significant runtime resources.

\lesson{\textit{Because of the health checks, hardware failures constitute a rare set of outcomes.}
Attributed hardware failures impact 19\% of GPU runtime and less than 1\% of jobs. This impact is significantly smaller once checkpointing is taken into account, which bounds lost work.}

\noindent{\textbf{Job-Level Failure Characterization.}} Attributed hardware failures can be broken down by attributed cause.
These causes can be further subdivided by server-level components, such as the GPU, the network, and various system components, such as the filesystem.
We show such GPU-hour normalized failure rates for \texttt{RSC-1}{}\arxivonly{\ and \texttt{RSC-2}{}} in Figure~\ref{fig:error-characterization}. \hpcaonly{We observe \texttt{RSC-2}{} follows similar trends.}
We attribute a failure to a cause if the cause was detected within the last 10 minutes or 5 minutes after a failing jobs lifetime (\texttt{FAILED} or \texttt{NODE\_FAIL}).
Note that we report the most likely cause of failure according to heuristics we developed indicating whether a node should be isolated for remediation. 
Some failures have multiple attributions (see Figure \ref{fig:error-characterization}).
Some \texttt{NODE\_FAIL} events are not associated with any health checks (c.f.,~\cite{cielo_memory}), likely because the node itself became unresponsive.
IB Links, filesystem mounts, GPU memory errors, and PCIe errors contribute heavily to the failure rates, however for IB Links in particular this seems to be dominated by a short period of many IB Link related job failures from a handful of nodes in the summer of 2024 as shown in Figure \ref{fig:failure_rate_over_time}. GSP Timeouts were caused by a code regression that was fixed with a driver patch (see Figure~\ref{fig:failure_rate_over_time}).

\hpcaonly{
\begin{figure}
\centering
\includegraphics[width=\columnwidth]{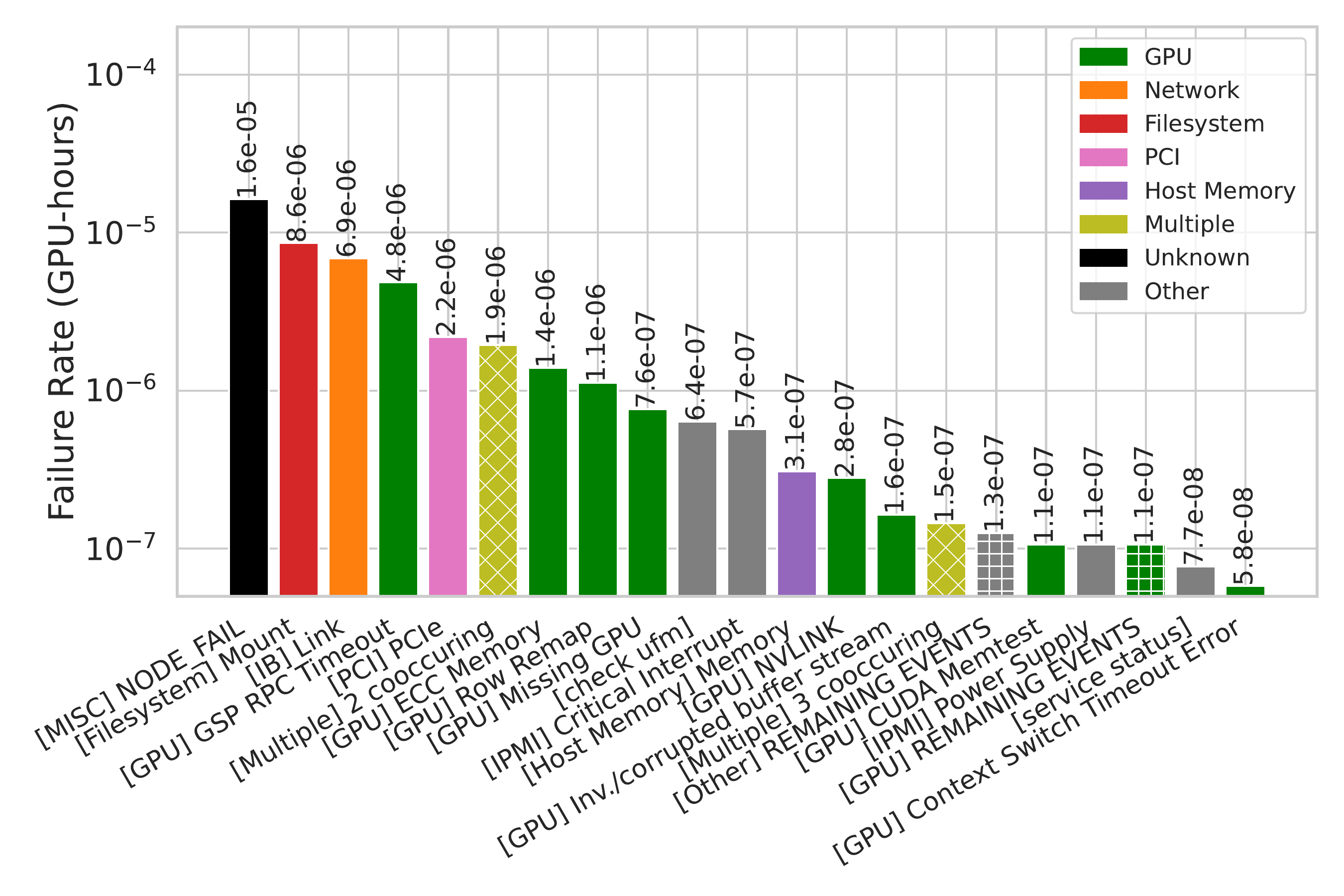}
\caption{Attributed hardware failures for all jobs on \texttt{RSC-1}{} expressed with per-GPU hourly rate.}
\label{fig:error-characterization}
\end{figure}
}
\arxivonly{
\begin{figure}
    \centering
    \begin{subfigure}[b]{\columnwidth}
    \ifthenelse{\boolean{fontcompat}}{{%
    \includegraphics[width=\columnwidth]
    {figures/compact_failure_cause_distribution_rsc_newf.pdf}%
    }}{{%
    \includegraphics[width=\columnwidth]
    {figures/compact_failure_cause_distribution_rsc_new.pdf}%
    }}
    \caption{\texttt{RSC-1}{}}
    \end{subfigure}
    \vspace{1em}

    \begin{subfigure}[b]{\columnwidth}
    \ifthenelse{\boolean{fontcompat}}{{%
    \includegraphics[width=\columnwidth]{figures/compact_failure_cause_distribution_rsc2_newf.pdf}%
    }}{{%
    \includegraphics[width=\columnwidth]{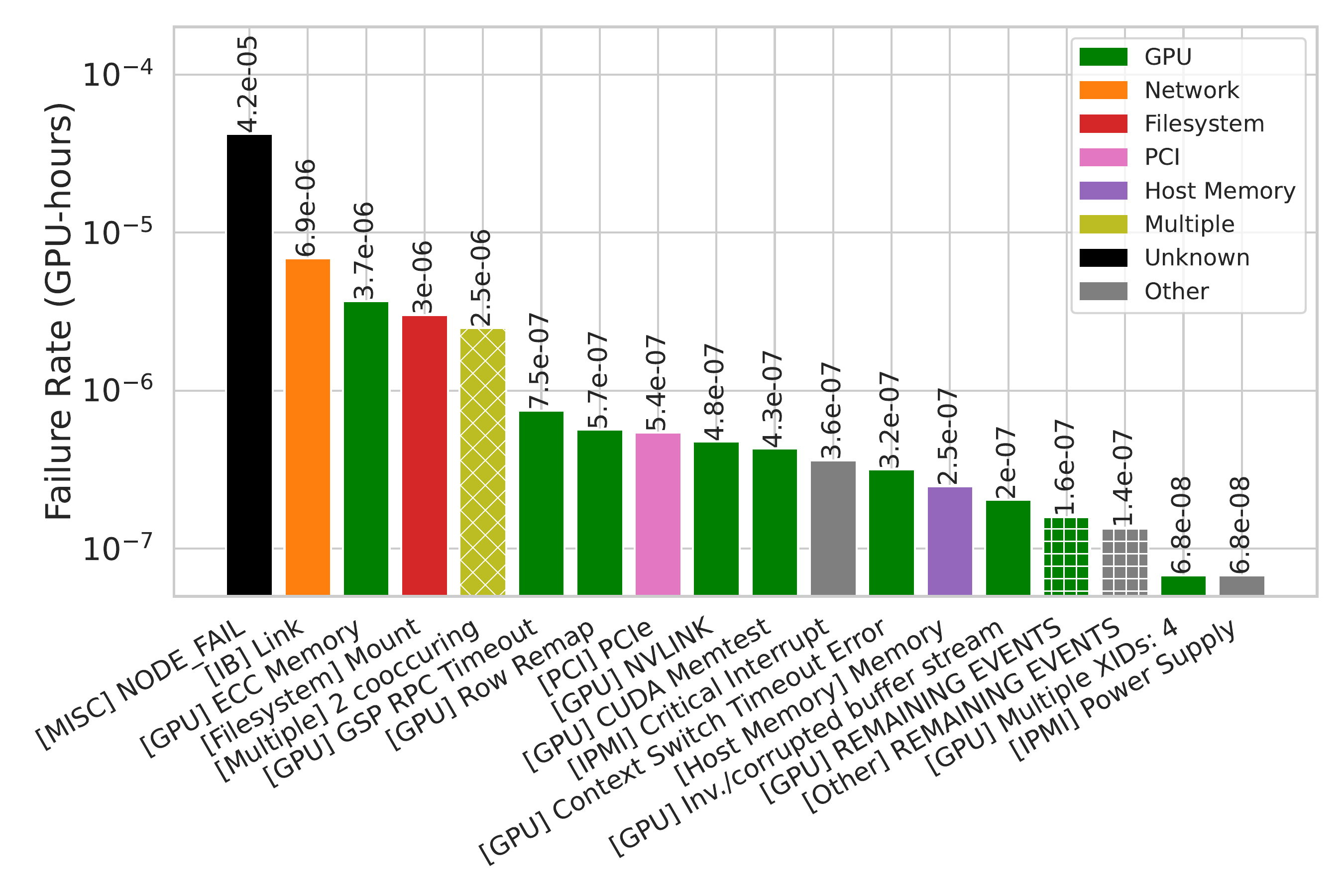}%
    }}
    \caption{\texttt{RSC-2}{}}
    \end{subfigure}
    \vspace{1em}

    \caption{Attributed hardware failures on \texttt{RSC-1}{} and \texttt{RSC-2}{} expressed with per-GPU hourly rate.}%
  \label{fig:error-characterization}%
\end{figure}
}

Failures may co-occur---3\% and 5\% of hardware failures on \texttt{RSC-1}{}/\texttt{RSC-2}{} have co-occuring events of similar priority.
For example, we observe PCIe errors often co-occur with XID 79 (GPU falling off the bus) and IPMI ``Critical Interrupt'' events.
On \texttt{RSC-1}{} (and \texttt{RSC-2}{)}, we observe 43\% (63\%) of PCI errors co-occur with XID 79 and 21\% (49\%) have all 3 event types.
This is expected, as all of these checks have overlap with PCIe and bus health.
Our data also appears to agree with decade-old studies on row-remapping, ECC errors, and falling off the bus~\cite{gpu_failures} being common, especially when considering that PCIe errors are highly correlated with XID 79.
We additionally observe that 2\% (and 6\%) of IBLink failures co-occur with GPU failures, such as falling off the bus, which may indicate a correlation with PCIe.

\lesson{\textit{Many hardware failures are unattributed, and the most common attributed failures are due to the backend network, the filesystem, and GPUs}.
GPUs show a rich error category due to fine-grained XIDs, though the top error codes are memory related.
PCIe bus errors and GPU falling off the bus are also common and are correlated.
CPU memory and host services are less likely to affect applications.
}

\vspace{0.25cm}
\noindent\textbf{Evolution of Failure Rate over Time.} We now turn our analysis to larger jobs, therefore switching to node-level (rather than GPU-level) analysis. In Figure~\ref{fig:failure_rate_over_time}, we show how failures manifest for \texttt{RSC-1}{} over the last year (plotting failure rates using a 30 day rolling average), illustrating:

\begin{itemize}[leftmargin=*]
\item \textbf{Failure rate is constantly changing.} We see periods where e.g., failure rate is $\sim$2.5 failures per 1000 node-days on \texttt{RSC-1}{} and periods where failure rate spikes as high as $\sim$17.5 failures per 1000 node-days (an order of magnitude higher). 
\item \textbf{Failure modes ebb and flow.} In late 2023, XID errors from a driver bug were the dominant source of job failures on \texttt{RSC-1}{}; this issue was resolved. In spring of 2024, after adding a new health check for mounts that were downing nodes, this became a key failure mode on \texttt{RSC-1}{}. In early summer of 2024, a spike of IB Link failures on a small number of offending nodes temporarily drove up the failure rate on both clusters.
\item \textbf{New health checks expose new failure modes.} We mark the time that new health checks were added to the cluster. The addition of a new health check, usually in response to an anecdotal report of a previously unchecked failure mode, has a tendency to cause an apparent increase in failure rate, simply because we are suddenly able to see a failure mode that was likely previously present. 
\end{itemize}

\lesson{\textit{Cluster failures are dynamic and reducing cluster failure rate is a continuous battle.} New workloads and software updates mean the cluster is constantly changing.}

\hpcaonly{
\begin{figure}
\centering
\includegraphics[width=\columnwidth]{figures/failure_rate_over_time_rsc_only.pdf}
\vspace{-15pt}
\caption{Evolution of cluster failure rate for \texttt{RSC-1}{} broken down by failure mode for 128 GPU+ jobs. Annotated vertical lines show the dates of introduction for various different health checks during the course of the year.}
\label{fig:failure_rate_over_time}
\vspace{-5pt}
\end{figure}
}
\arxivonly{
\begin{figure}
\centering
\ifthenelse{\boolean{fontcompat}}{{%
\includegraphics[width=\columnwidth]{figures/failure_rate_over_time_bothf.pdf}%
}}{{%
\includegraphics[width=\columnwidth]{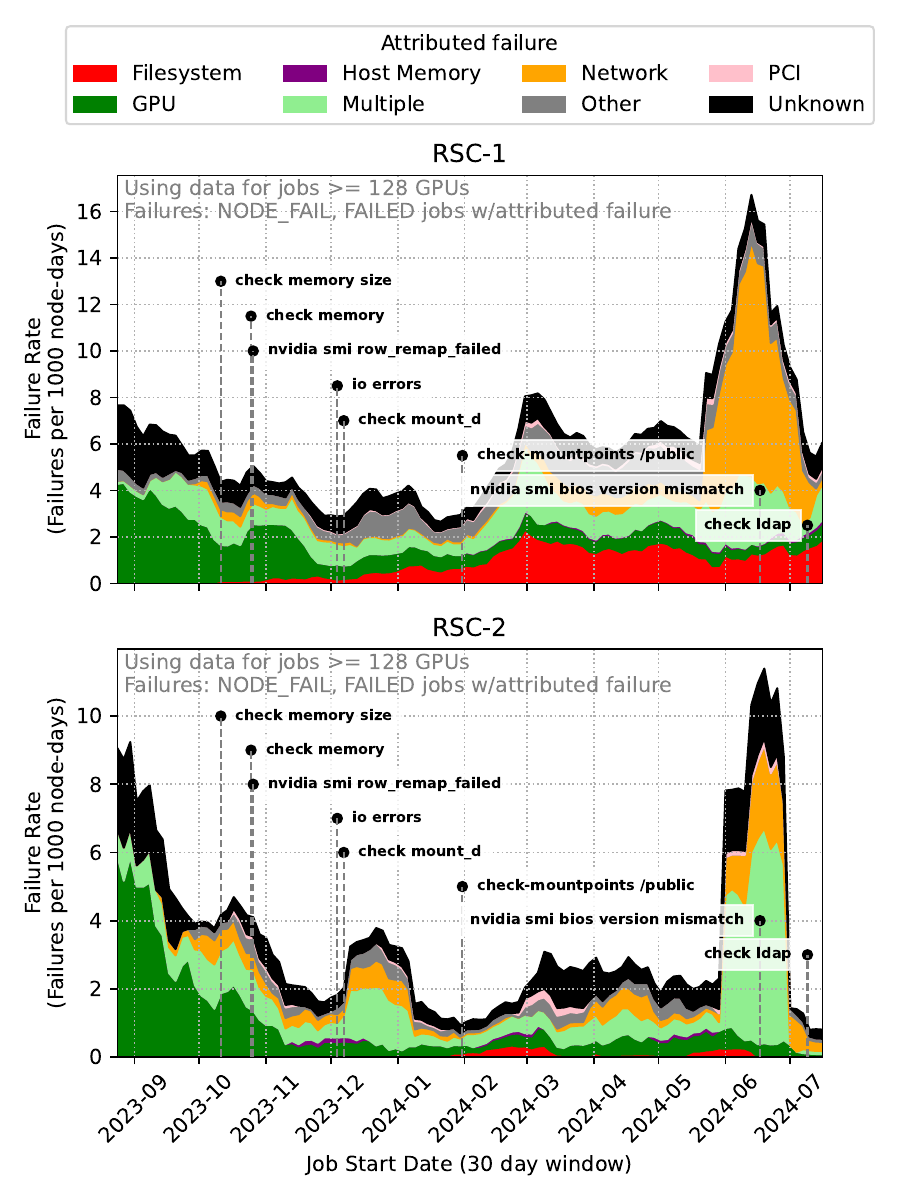}%
}}
\vspace{-15pt}
\caption{Evolution of cluster failure rate for \texttt{RSC-1}{} and \texttt{RSC-2}{} broken down by failure mode. Annotated vertical lines show the dates of introduction for various different health checks during the course of the year.}%
\label{fig:failure_rate_over_time}%
\vspace{-5pt}
\end{figure}
}

\vspace{0.25cm}
\noindent{\textbf{Training Job Diversity.} We have a diverse collection of training jobs in terms of job size and the overall consumed GPU hours.} The scheduler must consider job size diversity and the corresponding training time to balance between training time performance, fairness of individual training jobs, and overall cluster utilization.

Figure~\ref{fig:job-size-distribution-submitted} depicts the distribution of job size for the \texttt{RSC-1}{} cluster. 
More than 40\% of training jobs use a single GPU for development or for model evaluation. There are only a few large-scale training jobs, which utilize thousands of GPUs in the research clusters. 
In the same figure, we also illustrate the corresponding percentage of GPU time consumed by the jobs.
Despite many 1-GPU training jobs, more than 66\% and 52\% of the overall GPU time comes from 256+ GPU jobs for the \texttt{RSC-1}{} and \texttt{RSC-2}{} clusters, respectively.
Compared with production model training for LLaMa 3.0~\cite{llama3-model-card}, the diversity of training job sizes and the respective training time in the research clusters poses unique challenges to the design of an effective ML scheduler.

\lesson{\textit{Over 90\% of jobs are less than 1 server large, but represent less than 10\% of GPU time}. \texttt{RSC-1}{} tends to have more 8 GPU jobs compared to \texttt{RSC-2}{}, which tends to have 1 GPU jobs. \texttt{RSC-1}{} tends to have the largest jobs.}

\hpcaonly{
\begin{figure}
  \centering
  \includegraphics[width=\columnwidth]{figures/job_size_distribution_rsc_only.pdf}
  \caption{Job distribution by fraction of jobs and fraction of compute across \texttt{RSC-1}{}.}
  \label{fig:job-size-distribution-submitted}
\end{figure}
}

\arxivonly{
\begin{figure}
  \centering
  \ifthenelse{\boolean{fontcompat}}{{%
  \includegraphics[width=\columnwidth]{figures/job_size_distribution_bothf.pdf}%
  }}{{%
  \includegraphics[width=\columnwidth]{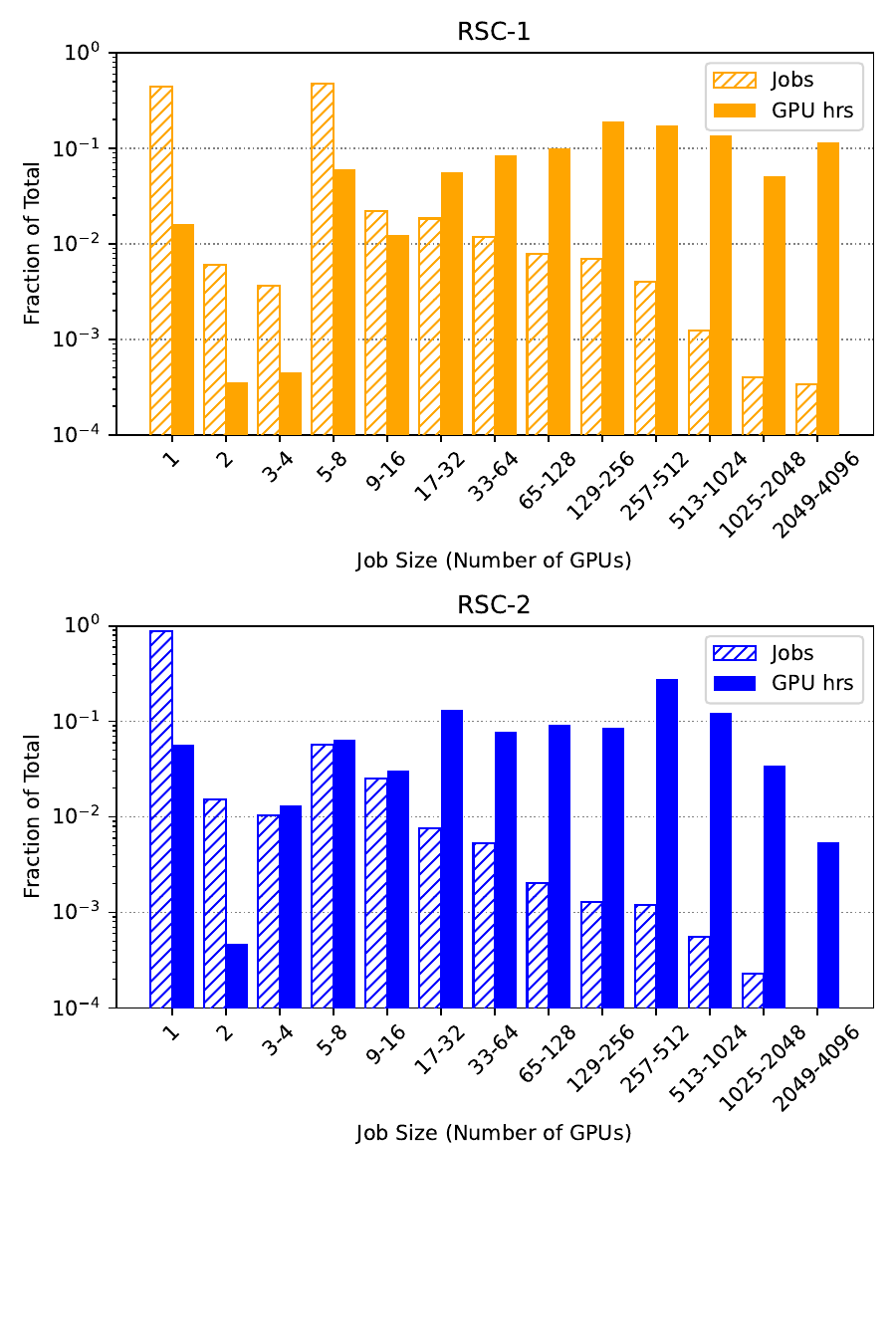}%
  }}
  \vspace{-5em}
  \caption{Job distribution by fraction of jobs and fraction of compute across \texttt{RSC-1}{} and \texttt{RSC-2}{}.}%
  \label{fig:job-size-distribution-submitted}%
\end{figure}
}

\begin{figure}
  \centering
  \ifthenelse{\boolean{fontcompat}}{{%
  \includegraphics[width=\columnwidth]{figures/mttf_by_job_sizef.pdf}%
  }}{{%
  \includegraphics[width=\columnwidth]{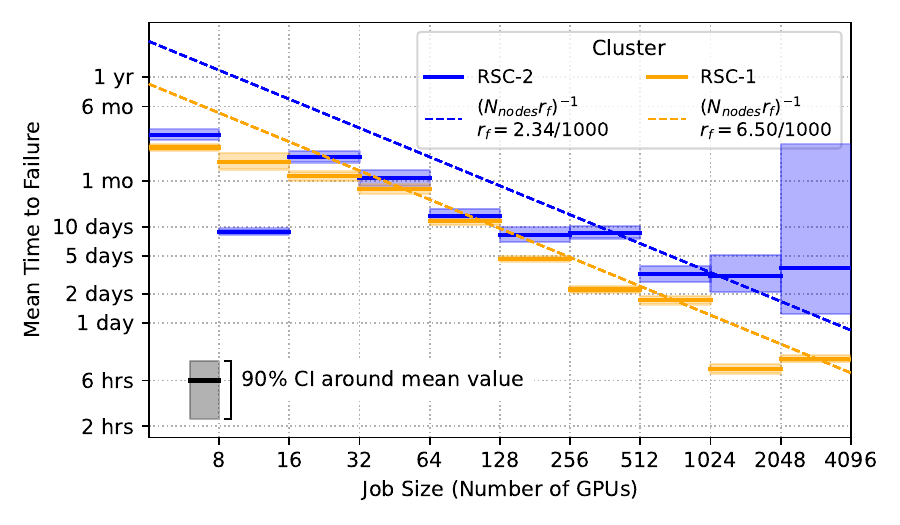}%
  }}
  \caption{MTTF analysis by job sizes for \texttt{RSC-1}{} and \texttt{RSC-2}{}, rounded up to the next multiple of 8 GPUs. \textbf{CI}: Confidence Interval (90\%). MTTF decreases predictably with scale.}%
  \label{fig:mttf-analysis}%
\end{figure}

\vspace{0.25cm}
\noindent{\textbf{MTTF Decreases at Scale.} Figure~\ref{fig:mttf-analysis} illustrates that the mean-time-to-failure (MTTF) of 1024-GPU jobs is 7.9 hours---roughly 2 orders-of-magnitude lower than 8-GPU jobs (47.7 days). As shown in \S\ref{sec:failure-taxonomy}, training failures stem from various factors, ranging from user programs to system software to hardware faults. 
Empirically, hardware reliability shrinks inversely proportional to the number of GPUs, with more consistent trends starting at 32 GPUs. 90\% confidence intervals are generated by fitting a Gamma distribution. 

We also show in Figure \ref{fig:mttf-analysis} that the theoretical expected MTTF ($\text{MTTF}\propto 1/N_{\text{gpus}}$) derived from cluster node failure rate: $\text{MTTF} = (N_{\text{nodes}} r_f)^{-1}$ where $r_f$ is calculated using total number of failures and node-days of runtime for all jobs $>128{}$ GPUs, matches well with observed MTTF numbers for larger jobs ($>$ 32 GPUs).

Based on the failure probability we observed from training jobs running in real-world research clusters at-scale, we project the MTTF for 16384 GPU jobs to be 1.8 hours and for 131072 GPU jobs to be 0.23 hours. 
To maximize \ettrabbr{} in the presence of failures (\S\ref{sec:metrics}), we must accelerate the process of \textit{failure detection} and \textit{recovery}. Taking a step further, making large model training fault-tolerant to failures is imperative to training productivity.
Note that for smaller jobs, we observe less predictable MTTFs, mostly due to experimental usage patterns that cause correlated \texttt{NODE\_FAIL}.
}

\lesson{\textit{While failures don't directly impact most jobs, large jobs are significantly impacted by failures, with failure rates matching theoretical trends.} Already at 4k scale, MTTF is around 10 hours and expected to decrease further at scale for \texttt{RSC-1}{}. MTTF projections closely match empirical MTTFs for 4 to 512 servers for \texttt{RSC-1}{}.
For \texttt{RSC-2}{}, the projection is similar, though the empirical MTTF data fluctuates more for 16 GPUs, partially due to a group of related jobs causing multiple \texttt{NODE\_FAIL}, and overall tends to be slightly more reliable than the \texttt{RSC-1}{} projected trend.
Some of this difference may be explained by different workloads triggering different failure causes e.g., Filesystem Mounts in Figure~\ref{fig:error-characterization}.
}

\vspace{0.25cm}
\noindent{\textbf{Preemptions and Failure Cascades.}
A second-order effect of job failures is their effect on other, lower priority and likely smaller jobs---resulting in \textit{cascades}~\cite{borg}.
In practice, large jobs tend to be higher priority jobs and small jobs are the lowest priority.
By virtue of being high priority, large jobs are scheduled quickly by preempting the low priority jobs.
When a large, high priority job fails due to hardware instability, Slurm is configured to reschedule it, possibly preempting hundreds of jobs in the process.
\arxivonly{The worst-case version of this is a \textit{crash loop}~\cite{google_trace_2012}, where a single job is configured to requeue on failures (e.g., by using exception handling in the submission script).}
In the period we observe, we see a 1024 GPU job \texttt{NODE\_FAIL} and subsequently requeue 35 times, causing a total of 548 preemptions (over 7k GPUs).
Such situations should be avoided, as they cause excessive churn in the cluster, resulting in lost goodput.

Preemptions are a second-order effect when considering job failures.
In our clusters, to help ensure even the lowest priority jobs are able to make progress, preemptions can only occur after two hours of runtime.
Nevertheless, without precise checkpointing, some work will be lost when job preemption occurs.
Critically, large jobs 1) are expected to lose significant work and 2) fail more frequently (Figure~\ref{fig:mttf-analysis}), resulting in quadratic goodput costs as a function of job size.
To estimate the impact of various sources of goodput loss on the overall cluster goodput, including preemptions occuring due to a rescheduled failed job, we assume that all jobs checkpoint hourly (we find this is a typical checkpoint interval for larger jobs on the RSC clusters), giving an average of half an hour of lost work.
Using the Slurm logs, we determine which jobs
\begin{enumerate*}[label=\protect\circled{\arabic*}]
    \item received a \texttt{NODE\_FAIL} (cluster-related issue) or a \texttt{FAILED} status that we attributed a hardware issue,
    \item were preempted (\texttt{PREEMPTED} status) because of an instigating \texttt{NODE\_FAIL} or \texttt{FAILED} job,
\end{enumerate*} 
and estimate the lost goodput (the minimum value of the jobs runtime and 30 minutes, multiplied by the number GPUs allocated to the job). 
Figure~\ref{fig:preemption_vs_failure} shows that, as expected, most lost goodput ($y$-axis) from failures and second-order preemptions (in terms of wasted runtime, ignoring a possibly large impact on resource fragmentation)  on \texttt{RSC-1}{} is due to large jobs at the scale of 2-4 thousand GPUs ($x$-axis). \arxivonly{On \texttt{RSC-2}{}, moderate-sized jobs make up a higher fraction of the goodput loss due to differences in job makeup (see Figure \ref{fig:job-size-distribution-submitted}). Absolute goodput loss for \texttt{RSC-2}{} is also an order of magnitude smaller than for \texttt{RSC-1}{}, a consequence of differences in job makeup and failure rate.}
While optimizing large jobs is clearly important, 16\% of the total lost goodput resulting from hardware failures on \texttt{RSC-1}{} is due to second-order preemptions, which come from jobs of much smaller sizes.
These results indicate that the cluster as a whole is impacted beyond the failures themselves.

\hpcaonly{
\begin{figure}
  \centering
  \includegraphics[width=\columnwidth]{figures/goodput_loss_rsc_only.pdf}
    \caption{Impact to cluster goodput on \texttt{RSC-1}{} from attributed failures and second-order preemption-from-requeue costs.
    }
  \label{fig:preemption_vs_failure}
\end{figure}
}

\arxivonly{
\begin{figure}
  \centering
  \ifthenelse{\boolean{fontcompat}}{{%
  \includegraphics[width=\columnwidth]{figures/goodput_loss_bothf.pdf}%
  }}{{%
  \includegraphics[width=\columnwidth]{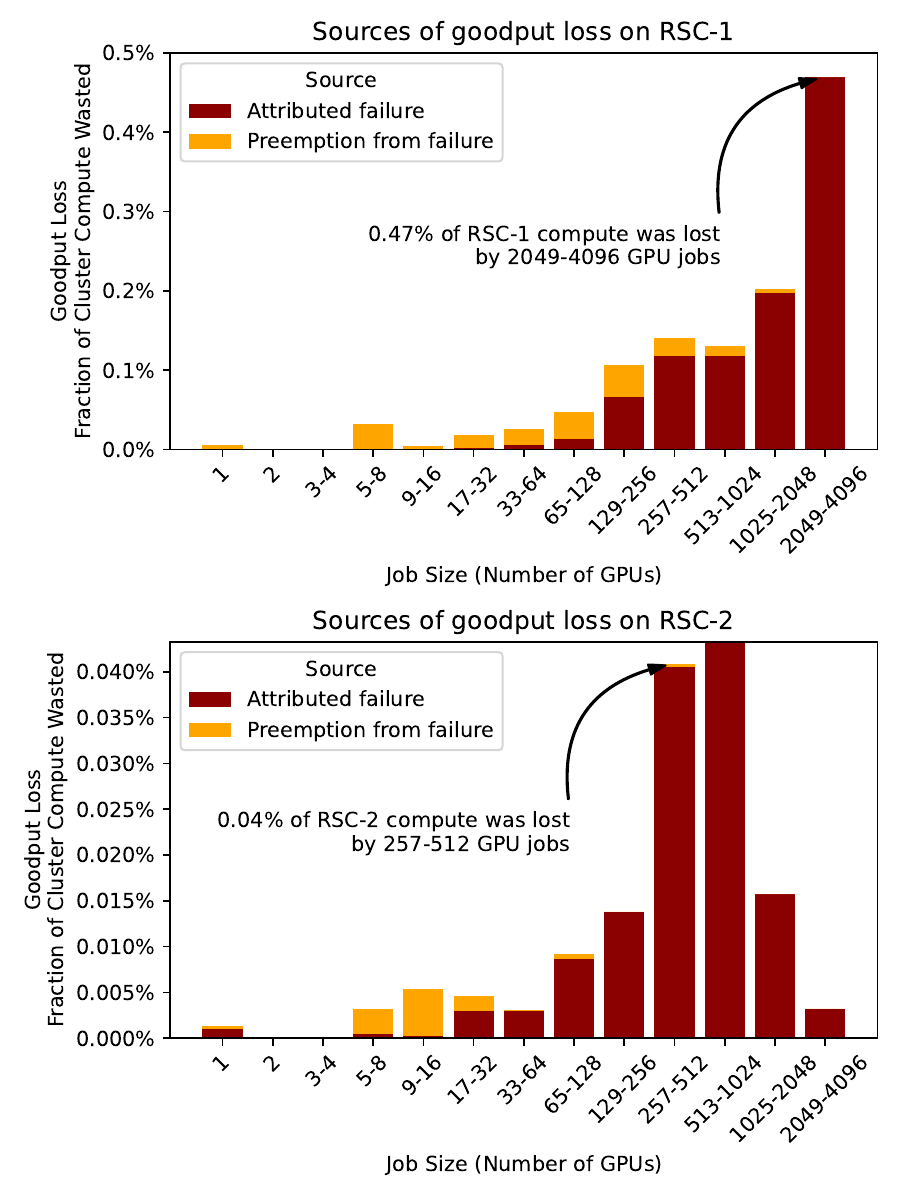}%
  }}
  \caption{Impact to cluster goodput on both \texttt{RSC-1}{} and \texttt{RSC-2}{} from attributed failures and  second-order preemption-from-requeue costs.}%
  \label{fig:preemption_vs_failure}%
\end{figure}
}
\lesson{\textit{Large, high priority jobs force scheduler churn upon failure.}
While first-order effects of a 1k+ GPU job failures are high, 16\% of total failure overhead comes from preempting other jobs.
The addition of job diversity therefore presents additional avenues for optimization.
}

\vspace{0.25cm}
\noindent{\textbf{Quantifying \ettrabbr{} at Scale.}} \ettrabbr{} provides an interpretable metric that quantifies the degree to which interruptions, queue time, and overhead impact training progress. Understanding how \ettrabbr{} scales with various quantities related to job configurations, scheduling, and failure statistics helps us understand the scale of the impact from various improvements.

In this section, we provide
\begin{enumerate*}[label=\protect\circled{\arabic*}]
\item an expected value formulation of \ettrabbr{} based on training job parameters, job priority, and cluster resource availability as inputs, and 
\item a design space exploration using job-level data to estimate \ettrabbr{} for the \texttt{RSC-1}{} and \texttt{RSC-2}{} clusters.
\end{enumerate*}
\begin{enumerate*}[label={For \protect\circled{\arabic*},}]
\item our formulation allows us to model a particular job's reliability properties by making 
assumptions about checkpoint frequency as well as checkpoint write and restart overhead.
Note that expected \ettrabbr{}, $\mathbb{E}[\text{\ettrabbr{}}]$, is most useful for longer training runs---by the law of large numbers, longer training runs will tend to have observed \ettrabbr{} values closer to this expectation.
Using our analytical formulation for expected \ettrabbr{}  helps us quickly estimate and understand the impact of optimizations, e.g., \textit{what is the impact of halving failure rate?}
\item
 we continue using the prior parameters as a tool for exploring the relative importance of different contributors to job overhead---exploring what the necessary requirements would be for reasonable \ettrabbr{} ($\sim$0.90) on the largest feasible \texttt{RSC-1}{} training runs (up to $\sim$2/3 of the cluster, or $\sim$12,000 GPUs) under the typical paradigm used today: checkpointing progress to disk and restoring upon a restart, without fault tolerant solutions like spare idle compute.
\end{enumerate*}

\textit{Approximating } $\mathbb{E}[\text{\ettrabbr{}}]$ \textit{ analytically}: 
First, define $Q$ as time the job was eligible to be scheduled but was waiting in the job queue, $R$ as productive runtime, and $U$ as unproductive runtime. The wallclock time, $W = Q + R + U$, is the total time since the job was first eligible to be scheduled until it completes. We consider the intervals between checkpoints as $\Delta t_{cp}$ with each checkpoint write incurring a constant time penalty of $w_{cp}$, the time it takes to perform initialization tasks (e.g., loading checkpoints, etc.) as $u_0$, and the expected queue time both after submission and after every interruption as $q$ %
(we assume queue times are drawn i.i.d.\ and are not systematically shorter after an interruption). The number of nodes the job consumes is $N_{\text{nodes}}$ and the cluster failure rate $r_f$ is the expected number of failures per node-day of runtime. The MTTF for the job is $(N_{\text{nodes}}r_f)^{-1}$.

The expected \ettrabbr{}, valid when $(u_0 + \Delta t_{cp}/2) \ll \text{MTTF} = (N_{\text{nodes}}r_f)^{-1}$, is

\begin{equation}
\mathbb{E}[\text{\ettrabbr{}}] \gtrsim \frac{1 - N_{\text{nodes}} r_f\left(u_0 + \frac{\Delta t_{cp}}{2}\right)}{1 + \frac{u_0 + q}{R} + \frac{w_{cp}}{\Delta t_{cp}} + N_{\text{nodes}} r_f q\left(1 + \frac{w_{cp}}{\Delta t_{cp}}  - \frac{\Delta t_{cp}}{2R}\right)}
\end{equation}

Which, for long-running, high priority jobs where queue time is much smaller than the MTTF and $R \gg q + u_0 + \frac{\Delta t_{cp}}{2}$ simplifies to 

\begin{equation}
\mathbb{E}[\text{\ettrabbr{}}] \approx \frac{1 - N_{\text{nodes}}r_f\left(u_0 + \frac{\Delta t_{cp}}{2}\right)}{1 + \frac{w_{cp}}{\Delta t_{cp}}}
\end{equation}

While a full derivation of the optimal checkpoint interval using the equation for \ettrabbr{} above is possible, a classic result by Daly and Young \cite{Daly_2006, Young_1974, Benoit_etal_2022, Bautista-Gomez_etal_2024} shows that, under some limiting assumptions, the optimal checkpointing interval is approximately

\begin{equation}
    \Delta t_{cp}^* = \sqrt{\frac{2w_{cp}}{N_{\mathrm{nodes}}r_f}}
\end{equation}

\ifthenelse{\boolean{isarxivversion}}{{%
\arxivonly{See Appendix \ref{appendix:\ettrabbr{}} for a more complete derivation of these results.}%
}}{{%
A more complete derivation of these results is presented in an extended report~\cite{kokolis2024revisitingreliabilitylargescalemachine}. 
}}

For the RSC clusters, $r_f \approx 5\times 10^{-3}$ failures per GPU node-day of runtime, $w_{cp} \approx 5~\mathrm{mins}$, $u_0 \approx 5-20~\mathrm{mins}$, and $(N_{\text{nodes}}r_f)^{-1} \gtrsim 0.1~\mathrm{day}$. Comparing to a Monte Carlo approach for computing the various expectations involved, even for large, long-running hypothetical jobs (e.g., 8k GPUs), we find that the approximation above is accurate to within $\sim$5\%.

\textit{Comparing to actual job run{}s}: We compare the expected value formulation of \ettrabbr{} above with observations of actual job runs observed on both clusters. A \textit{job run{}} is a collection of jobs (some may have different Job IDs) that are part of the same training task. We assume $\Delta t_{cp}$ is Daly-Young optimal, and that $u_0$ and $w_{cp}$ are both 5 minutes.
We focus on longer job run{}s with at least 48 hours of total training time and jobs that run with the highest priority.
Note that in calculating job run{} \ettrabbr{}, we do not consider health checks; we assume every job in the job run{} that does not exit cleanly (with a \texttt{COMPLETED} state assigned) is interrupted by an infra failure, whether it was caught or not, and this means that our data estimate of \ettrabbr{} should be an \emph{underestimate}.

\hpcaonly{
\begin{figure}
\centering
\includegraphics[width=0.9\columnwidth]{figures/ettr_vs_job_size_P0_rsc_only}
\caption{Comparing expected \ettrabbr{} \highlightchange{($\mathbb{E}[\ettrabbr{}]$)} from aggregate cluster and partition-level statistics \highlightchange{on both wait time and failure rate} with average estimated \ettrabbr{} from actual job runs, \highlightchange{assuming Daly-Young optimal checkpointing with 5 minute restart overhead and 5 minute checkpoint write overhead}. Error bars shown are \highlightchange{90\% CI} around the mean value of job run \ettrabbr{}. \highlightchange{\textbf{CI}: Confidence Interval is shown on empirical data.}}
\label{fig:ettr_vs_job_size}
\end{figure}
}

\arxivonly{
\begin{figure}
\centering
\ifthenelse{\boolean{fontcompat}}{{%
\includegraphics[width=0.9\columnwidth]{figures/ettr_vs_job_size_P0_bothf.pdf}%
}}{{%
\includegraphics[width=0.9\columnwidth]{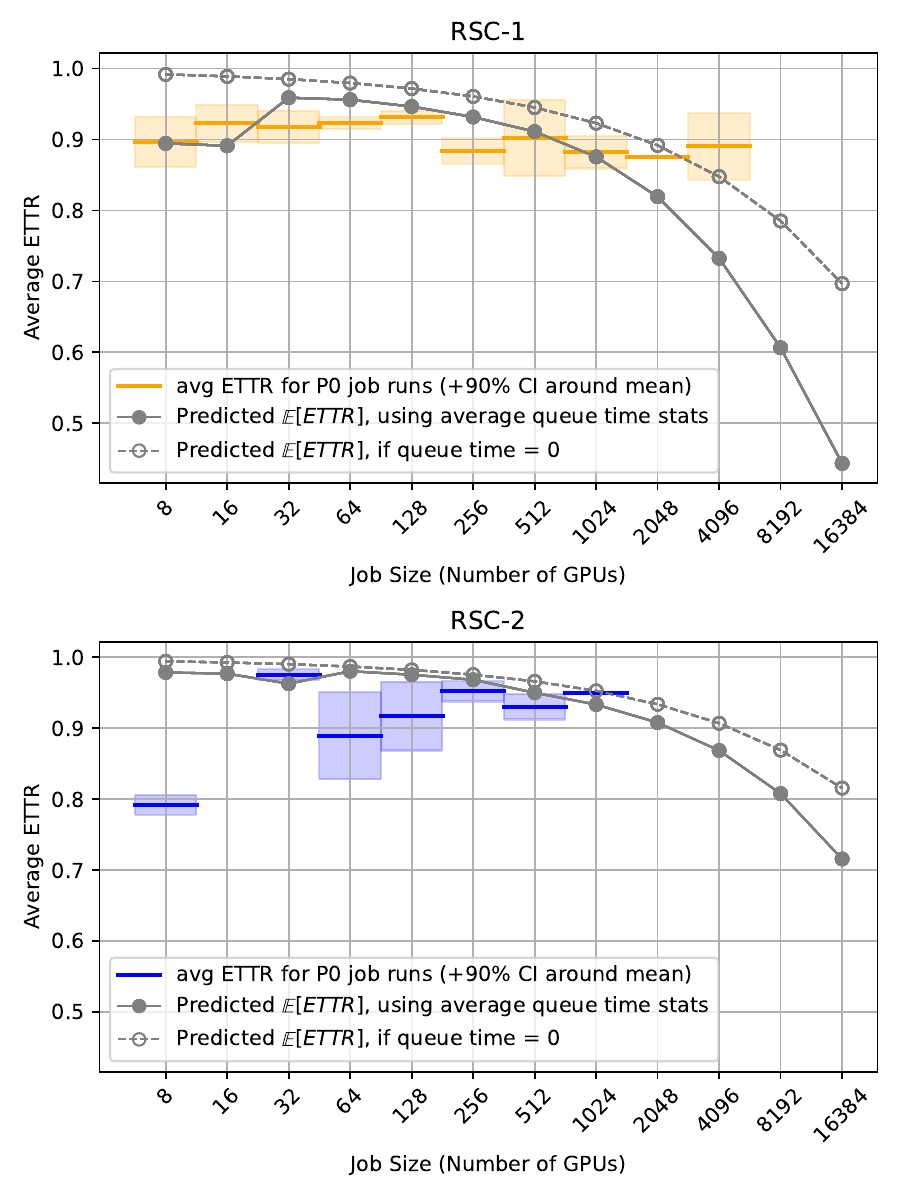}%
}}
\caption{Comparing expected \ettrabbr{} ($\mathbb{E}[\text{\ettrabbr{}}]$) from aggregate cluster and partition-level statistics on both wait time and failure rate with average estimated \ettrabbr{} from actual job runs, assuming Daly-Young optimal checkpointing with 5 minute restart overhead and 5 minute checkpoint write overhead. Error bars shown are 90\% CI around the mean value of job run \ettrabbr{}. \textbf{CI}: Confidence Interval is shown on empirical data.
}%
\label{fig:ettr_vs_job_size}%
\end{figure}
}

To obtain cluster-level failure rates $r_f$ needed to compute $\mathbb{E}[\text{\ettrabbr{}}]$, we count failures as taking all jobs (not just job runs) that use more than 128{} GPUs that are assigned a \texttt{NODE\_FAIL} status plus the number of jobs with a status of \texttt{FAILED} for which we can attribute a critical health check firing in the last 10 minutes of the job (or 5 minutes after completion). We then divide the number of failures by the number of node-days of runtime (the sum of runtime in days multiplied by the number of allocated nodes). We find nominally that \texttt{RSC-1}{} has an $r_f$ of 6.50{} failures per thousand node-days and \texttt{RSC-2}{} has a significantly lower $r_f$ of 2.34{} failures per thousand node-days. 
\arxivonly{This finding is also corroborated by looking at the rate at which GPUs are swapped in the cluster---we find \texttt{RSC-1}{} GPUs are swapped at $\sim$3 times the rate compared to \texttt{RSC-2}{}; both the GPU swap rate and failure rate differences may be due to differing workloads that tax GPUs on \texttt{RSC-1}{} more heavily.}

\vspace{0.25cm}
\noindent{\textbf{Analysis of \ettrabbr{} Results.}} Figure \ref{fig:ettr_vs_job_size} shows our findings. Our predictions of $\mathbb{E}[\text{\ettrabbr{}}]$ and average measured job run{} \ettrabbr{} agree fairly well, with measured job run{} \ettrabbr{} being generally smaller than predicted due to our conservative assumption that every state besides \texttt{COMPLETED} indicates an infra-related interruption. On \texttt{RSC-1}{}, the largest job runs ($> 1024$ GPUs) have systematically higher \ettrabbr{} than predicted by $\mathbb{E}[\text{\ettrabbr{}}]$. This is due to actual wait times for these larger job run{}s being shorter than average, possibly due to Slurm scheduling configurations that prefer larger jobs.

\textit{Looking towards the future}: The largest jobs on \texttt{RSC-1}{} today consume roughly a quarter of the cluster (4096 GPUs). High-priority research efforts may feasibly consume up to 2/3 of the cluster for a short amount of time. Figure \ref{fig:largerun_ettr} shows projected \ettrabbr{} as a function of both failure rate and checkpoint write overhead. To obtain a good \ettrabbr{} ($\gtrsim 0.9$) for a job of this size, \texttt{RSC-1}{} failure rate either needs to improve from 6.50{} to $\sim$1 or checkpoint write overhead needs to be under a minute $\mathcal{O}(10 \mathrm{s})$, achievable with asynchronous checkpoint writing strategies~\cite{Gemini_Wang2023},
though note that our analytical model was created with a classical checkpoint strategy in mind.

\begin{figure}
\centering
\ifthenelse{\boolean{fontcompat}}{{%
\includegraphics[width=\columnwidth]{figures/100k_ettrf.pdf}%
}}{{%
\includegraphics[width=\columnwidth]{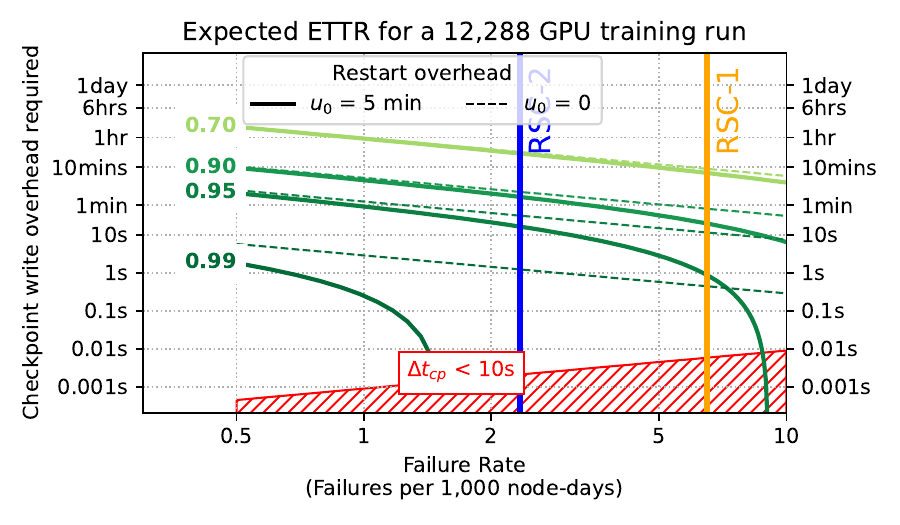}%
}}
\caption{Checkpoint and failure rate requirements for large 12k GPU-scale job runs.
Contours interpolate between poor ETTR (0.7, light green) to almost perfect ETTR (0.99, dark green) as a function of cluster failure rate and checkpoint write overhead.
Checkpoint intervals smaller than 10s shown in red.
}
\label{fig:largerun_ettr}
\end{figure}

\lesson{\textit{The RSC clusters are highly efficient (\ettrabbr{} 0.85-0.9) for their largest and highest priority jobs.} 2+ day 2048-4096 GPU job runs on \texttt{RSC-1}{} show an average \ettrabbr{} of around 0.9 assuming checkpoint intervals are Daly-Young optimal and that both restart overhead and checkpoint write overhead is 5 minutes, despite both clusters being congested and shared resources with 2-hour minimum time-to-preemption requirements for even the lowest priority jobs. To reach \ettrabbr{} of 0.9 for the largest feasible training runs on \texttt{RSC-1}{} ($\sim$12,000 GPUs, taking two thirds of \texttt{RSC-1}{}), checkpoint write time overhead needs to be on the order of $\sim$10 seconds or failure rate needs to dramatically improve from 6.50{} to $\sim$1.}

\section{Improving Cluster Reliability and Efficiency}%
\label{sec:cluster-optimization}
This section discusses mitigations that we have put in place to increase cluster reliability.
Health checks are but one piece of the mitigations, and they are especially effective at finding random node failures.
However, failures may be correlated to a particular node what we call a \textit{lemon node}, due to a misconfiguration, aging, or hardware defects.
Therefore, health check infrastructure can be generalized to find recurring problematic nodes (\S\ref{sec:lemon-detection}).
We additionally extend our practice outside the server itself by including network level mitigations for when routes in the network become unreliable (\S\ref{sec:adaptive_routing}).
Lastly, we outline how cluster-level metrics can be influenced by cluster design and the workload itself.

\subsection{Identifying and Repairing Lemon Nodes}
\label{sec:lemon-detection}
While health checks provided initial protection from multiple jobs failing due to the same failure, in practice, we observed that certain nodes had above-average rates of job failures.
Because the rates of failure were correlated with a specific node, we suspected that the hardware was degrading or the node was running misconfigured software.
Unfortunately, it is difficult to find such nodes quickly, because it requires observing their failure behavior over a long period of time to obtain statistical significance. Worse yet, such nodes keep on attracting new jobs upon failures, only to fail them eventually and bring down overall \ettrabbr{}.
This section outlines how we setup such a detection pipeline.

In the presence of faulty nodes, either due to transient or permanent faults from different hardware components of the training cluster, researchers would manually exclude nodes that cause job failures based on past experience. This practice is, however, not scalable and aggressive exclusion of nodes may lead to capacity starvation. 

To improve the effective training time ratio, we design \textit{\textbf{lemon detection}} to proactively identify, isolate, and replace  \textit{\textbf{lemon nodes}} from the machine learning job scheduler (i.e., Slurm). Lemon nodes are the servers that cause repeating job failures but cannot be identified by existing mechanisms like health checks and repair workflows. As what \S\ref{sec:analysis} previously shows, one of the most important factor causing training job failures is {\tt NODE\_FAIL}, stressing the importance of proactively handling lemon nodes. 

\textbf{Lemon Detection Signals.}
Among tens of detection signals available on each node, the following ones correlate with lemon nodes the most:
\begin{enumerate*}[label=\protect\circled{\arabic*}]
    \item \arxivonly{{\tt excl\_jobid\_count}:} Number of distinct jobs that excluded a node.
    \item \arxivonly{{\tt xid\_cnt}:} Number of unique XID errors a node experienced.
    \item \arxivonly{{\tt tickets}:} Count of repair tickets created for a node.
    \item \arxivonly{{\tt out\_count}:} Number of times node was taken out of availability from the scheduler.
    \item \arxivonly{{\tt multi\_node\_node\_fails}:} Number of multi-node job failures caused by a node.
    \item \arxivonly{{\tt single\_node\_node\_fails}:} Number of single-node job failures caused by a node.
    \item \arxivonly{{\tt single\_node\_node\_failure\_rate}:} Rate of single-node job failures on a node.
\end{enumerate*}
We can view these signals as potential features into a binary classification model, though the results we report were tuned manually based on accuracy and false positive rate of predicted lemon nodes.

\arxivonly{
\begin{figure}
\centering
\ifthenelse{\boolean{fontcompat}}{{%
\includegraphics[width=\columnwidth]{figures/lemon_cdff.pdf}%
}}{{%
\includegraphics[width=\columnwidth]{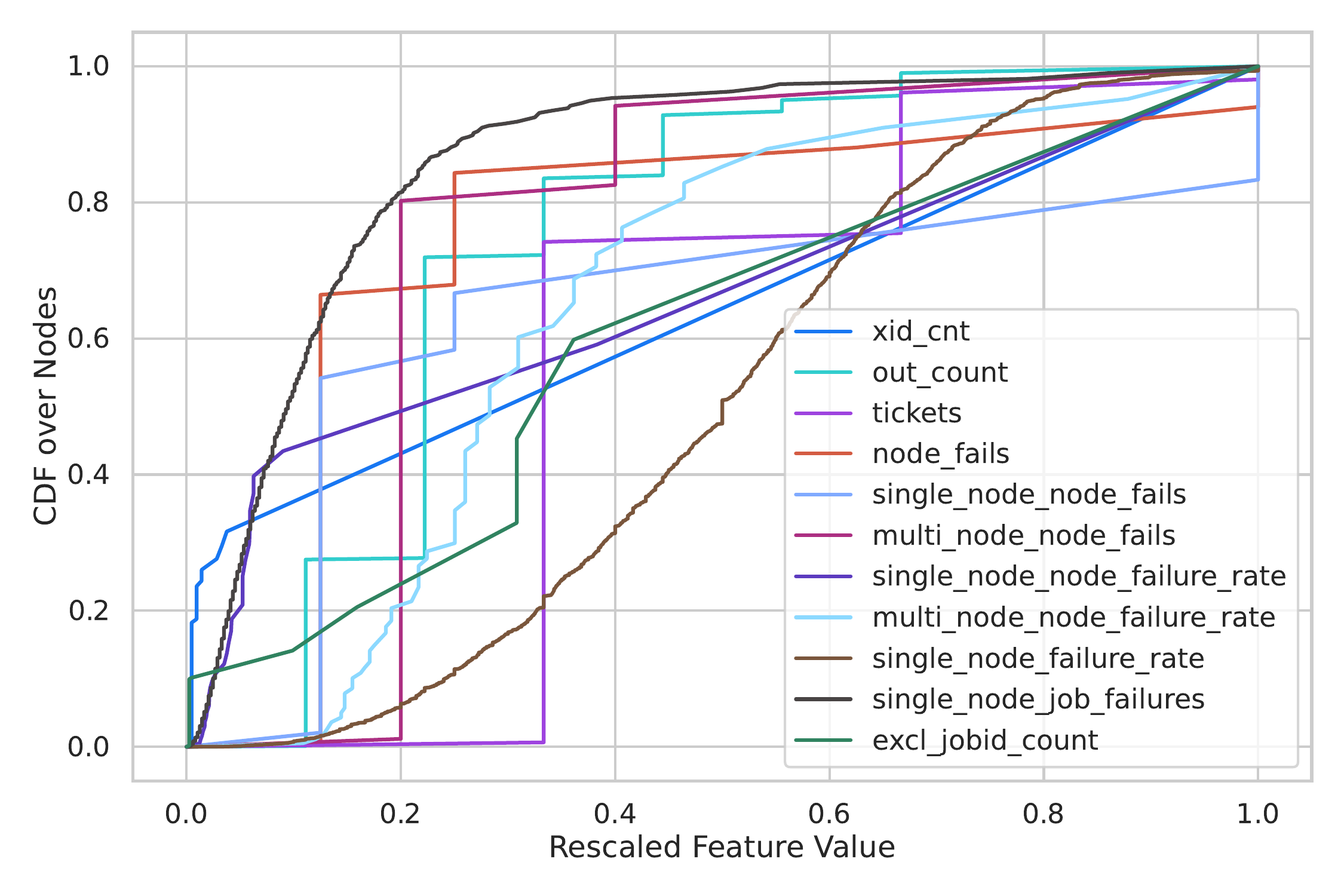}%
}}
\caption{Lemon Nodes Features compared to CDF of Nodes covering 28 days. Most features outside of failure data are highly sparse, resulting in non-smooth CDF behavior.}
\label{fig:lemon-node-cdf}
\end{figure}

Figure~\ref{fig:lemon-node-cdf} illustrates the distribution of the  signals based on a 28-day data snapshot for \texttt{RSC-1}{}, which we use to set the thresholds for detection criteria.
The x-axis represents the number of occurrence for the signal per GPU node, normalized from 0 to 1.
The y-axis represents the cumulative and normalized number of GPU nodes that experienced each signal. We found that user reported signal on {\tt excl\_jobid\_count} did not have a strong correlation with node failures, yet a large number of nodes were excluded by at least one job. This motivates us to proactively detect lemon nodes instead of leaving the burden of lemon detection to ML developers.}

\begin{table}
\centering
\addtolength{\tabcolsep}{-0.4em}
\begin{tabular}{lllllllll}
\toprule
Optics & CPU & PSU & NIC & EUD & PCIE & DIMM &  GPU & BIOS \\
\midrule
2.6\% & 2.6\% & 5.1\% & 7.7\% & 10.3\% & 15.4\% & 20.5\%  &  28.2\% & 7.7\% \\
\bottomrule
\end{tabular}
\caption{Fraction of Lemon Node Root Causes}
\label{tab:lemon-node-detection}
\end{table}

More than 85\% of identified lemon nodes failed one of the tests.
The failures are classified in Table~\ref{tab:lemon-node-detection}.
We designed, implemented, and evaluated the lemon detection mechanism to successfully identify 40 faulty nodes across \texttt{RSC-1}{} (24 nodes) and \texttt{RSC-2}{} (16 nodes), achieving more than 85\% accuracy. The identified lemon nodes represent 1.2\% of \texttt{RSC-1}{}'s footprint, 13\% of daily jobs, and 1.7\% of \texttt{RSC-2}{}'s footprint. Our lemon node detection mechanism led to 10\% reduction in large job failures (512+ GPUs), from 14\% to 4\%.

\lesson{\textit{Historic data is necessary to find defective nodes.} Implementing lemon node detection can improve large job completion rate by over 30\%.}

\subsection{Making Network Fabric Resilient via Adaptive Routing}
\label{sec:adaptive_routing}

The failure characterization analysis illustrates the significance of failures caused by the Infiniband link errors.
Just as servers may fail in various stages, for transient or permanently degraded components, network links may undergo similar behavior due to changes in physical or electrical properties.
Highly parallel model training at-scale will inevitably encounter faulty network links. 
These can be links with high error rates, flapping behavior that transitions between \texttt{up} and \texttt{down} states, permanently \texttt{down} links, or high congestion in multi-tenant environments. All of these can result in degraded performance for communication across the fabric.

Physical link replacement at scale is cumbersome. Thus, Infiniband fabrics come with switch-level techniques for tolerating link issues.
One such self-healing feature, called SHIELD~\cite{shield}, allows switches to coordinate around failed links.
However, even with such a feature enabled, the threshold for counting a link as down may be too conservative, resulting in re-transmissions at the protocol level along with possible network degradation.
In particular, in the bring-up phase of \texttt{RSC-1}{}, we observed as much as 50-75\% bandwidth loss.

Another, more advanced, feature is Adaptive Routing (AR) that dynamically adjusts routing decisions based on real-time network conditions. 
AR balances the traffic across all network links and increases aggregate link utilization. 
By allowing packets to avoid congested areas and unhealthy links, adaptive routing improves network resource utilization and efficiency. As such performance variance in training job performance attributed to network issues decreases. We have AR enabled in our clusters to increase performance predictability.

To showcase the importance of AR in our clusters, we performed two experiments.
In the first one, we introduced Bit Errors (BER) in our fabric using the \texttt{mlxreg} tool to modify the port registers in the fabric. Then, we run \texttt{All-Reduce} benchmark from NCCL-Tests~\cite{nccl_tests} with and without AR enabled across 512-GPUs.
The results in Figure \ref{fig:ar-bandwidth-linkerrors} show that AR is able to maintain much higher bandwidth under link errors.
In the second one, we simultaneously run multiple iterations of \texttt{All-Reduce} NCCL-Tests across 64 nodes in groups of two nodes (16 GPUs), to show how AR behaves under contention. 
Figure~\ref{fig:ar-bandwidth} shows that when flooding the fabric with multiple NCCL rings the performance variation is lower when using AR, and AR can achieve higher performance. 
This is because AR can shield GPUs from being bottlenecked by congested links. 
The impact of bad links in the fabric is spread across jobs as opposed to penalizing training jobs that happen to be mapped to the bad link(s) by enabling switches to select output port based on the port's load.

\lesson{\textit{The network must remove and route around failures.} Without resilience mechanisms in place, over 50\% of bandwidth may be lost.}

\begin{figure}[t]
        \centering
        \begin{subfigure}[b]{.49\linewidth}
            \ifthenelse{\boolean{fontcompat}}{{%
            \includegraphics[width=\linewidth]{figures/ar_bandwidth_plot_linkerrors_smallf.pdf}%
            }}{{%
            \includegraphics[width=\linewidth]{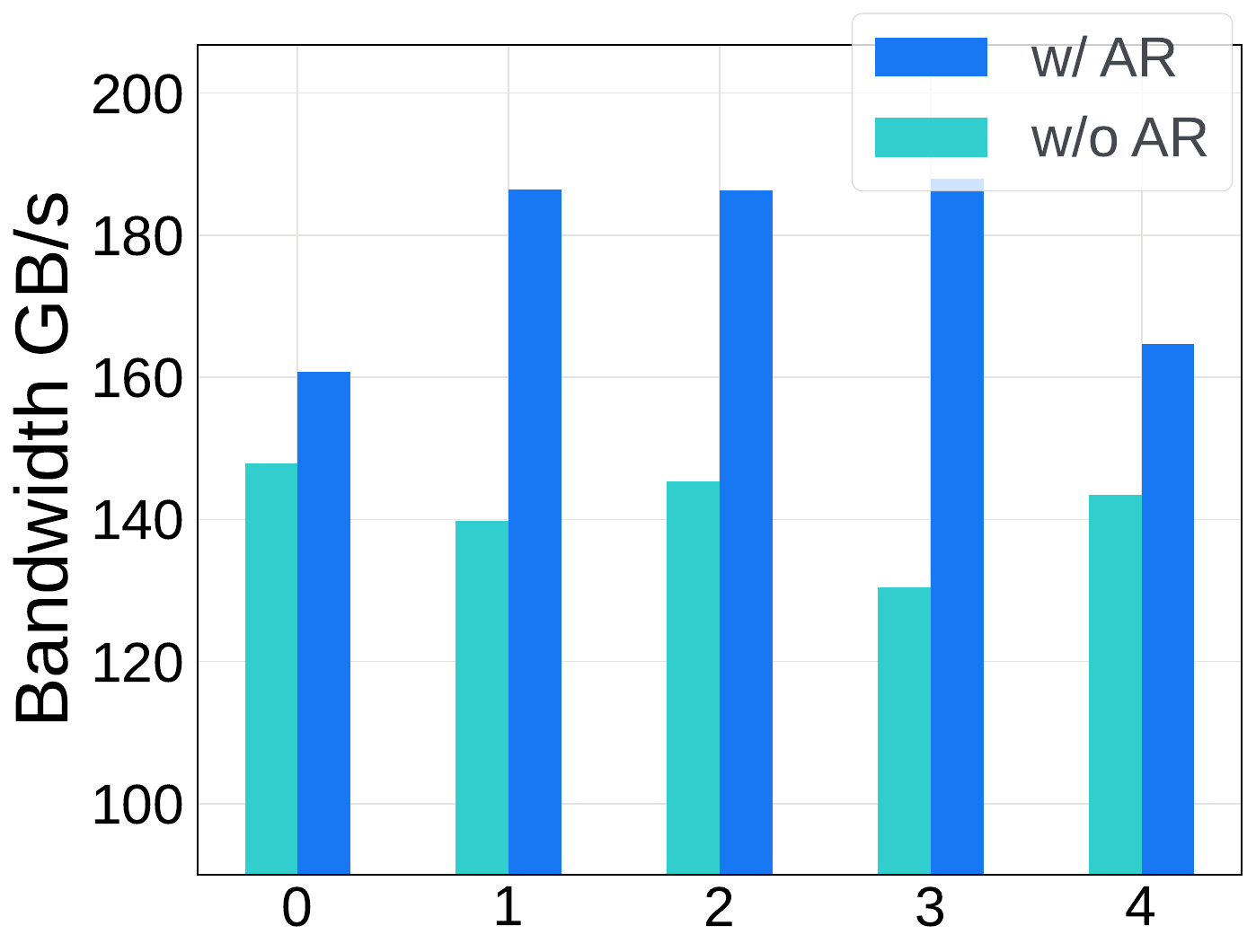}%
            }}
        \subcaption{}
        \label{fig:ar-bandwidth-linkerrors}
        \end{subfigure}
        \begin{subfigure}[b]{.45\linewidth}
            \ifthenelse{\boolean{fontcompat}}{{%
            \includegraphics[width=\linewidth]{figures/ar_bandwidth_plot_smallf.pdf}%
            }}{{%
            \includegraphics[width=\linewidth]{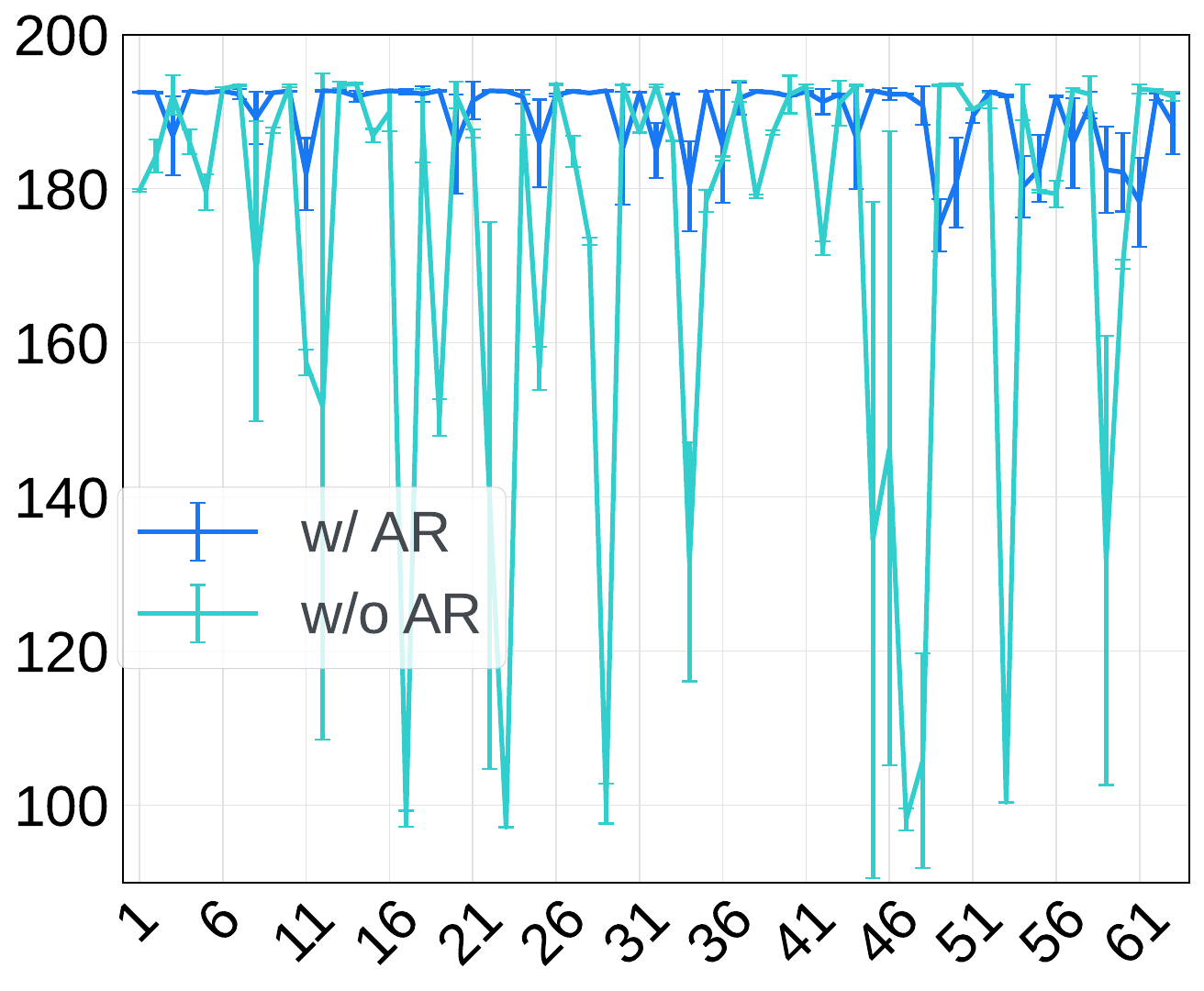}%
            }}
            \subcaption{}
            \label {fig:ar-bandwidth}
        \end{subfigure}
        \vspace{5pt}
            \caption {Bandwidth with and without AR for: (a) Five iterations of 512 GPU NCCL \texttt{All-Reduce} under link-errors. (b) 64 groups of 16 GPU NCCL \texttt{All-Reduce}.}
\end{figure}

\section{Key Lessons and Research Opportunities}%
\label{sec:opportunities}

This section summarizes lessons learned and highlight key opportunities to improve reliability, manage  infrastructure complexity, and co-design solutions.

    \noindent \textbf{Training Reliability:}
    As suggested by \textbf{Observations 1}, \textbf{2}, and \textbf{4}, keeping nodes and therefore jobs healthy is a major challenge.
    By \textbf{Observations 7}, \textbf{8}, \textbf{9}, and \textbf{10}, the impact of unhealthy infrastructure is felt asymmetrically across job sizes and priorities due to job-size dependent failure rates along with scheduling dynamics.
    We present how we run health checks and historic analysis to find transient failures, which inhibit a reliable compute substrate, as well as lemon nodes, whose removal can improve large job completion rates by over 30\% (\textbf{Observation 11}). 
    Our analysis does not focus on Silent Data Corruption (SDC) errors~\cite{SDC-meta,SDC-google}.
    A common technique to guard against SDCs is to monitor abrupt changes in gradients or activations during model convergence~\cite{understanding_hwfailure_sdc,drdna}.
    While understanding SDCs is important future work, our own anecdotal experience with operating a research clusters is that such events rarely warrant user complaints.
    
    Looking forward, we see significant opportunities in further exposing reliability information to the scheduler as well as distributed algorithms~\cite{hsia-isca2024,gpu_age_aware_scheduling}, such that work is partitioned to maximize reliability or goodput.
    Additionally, we note potential improvement in the network fabric itself in being resilient by, for example, being able to reconfigure its topology to route around failures, similar in spirit to what we present on Adaptive Routing (\S\ref{sec:adaptive_routing}, \textbf{Observation 12}). We therefore envision future infrastructure systems that attempt to make unreliability less noticeable rather than attempting to remove it altogether.
    We believe rethinking entire systems may be necessary when future GPU systems, such as the NVIDIA GB200~\cite{GB200}, will change the unit of repair from a server to a rack, creating incentives to avoiding downtime by coping with failure.

    Closer to the application side, recall that the \ettrabbr{} of a training job is a function of latency overheads associated with failures. An effective way to improve \ettrabbr{} is to minimize checkpoint~\cite{Gemini_Wang2023} and restart latency costs, while also reducing the probability of restarts due to failures.
    As observed in prior work~\cite{megascale}, certain operations, such as NCCL initialization, can scale poorly with the number of GPU nodes. Looking forward, it is, therefore, important for future software systems to support fast and reliable routines---optimizing the latency cost of restarts. We expect optimizations, such as, replacing MPI-like collectives entirely and making preflight hardware tests more efficient, as key future avenues.

    \noindent \textbf{Debugging Tools:}
    Once many significant failures are eliminated via health checks, the remaining failures often manifest with proximal failures that do not immediately suggest a root cause, as suggested by \textbf{Observations 3}, \textbf{5}, and \textbf{6}. NCCL timeouts are one of the most common symptoms of a failure whose root cause could be network infrastructure, buggy model code, or other stuck components.
    Regularly checking hardware infrastructure health (\S\ref{sec:health-checks}) reduces the frequency of NCCL timeouts by finding faulty network or hardware issues proactively before they manifest as NCCL kernels becoming stuck. Identifying the remaining root causes of NCCL timeouts may require new health checks and tools.

    We can improve the success rate of training runs by retroactively identifying the root cause of a NCCL timeout, by comparing logged data across different ranks participating in the collective. By logging which ranks started each collective and the dependencies between collectives, we can find the first collective where some ranks started the collective but others did not, and we can further investigate the missing ranks. If all ranks entered but did not leave a collective, we can examine the network traffic within the collective to identify which component did not send or receive an expected message~\cite{flightrecorder}. Removing the culprit rank or network components from future runs will reduce the likelihood of those runs hitting the same issue. Better diagnosis and debugging tools are needed for efficient, yet reliable large-scale distributed training.
    One can imagine extending existing management tools, such as IPMI~\cite{ipmi}, to deliver machine debug information in an out-of-band network and closing the attribution gap.

    \noindent \textbf{New Programming Models and Algorithms:}
    This work primarily focuses on traditional SPMD programming models with traditional Bulk-Synchronous Parallel~\cite{Valiant_BSP} semantics.
    Going forward, relaxing both of these constraints may unlock additional reliability, since such future programs may be able to more gracefully tolerate failures.

    On the system side, one confounding factor for diagnosing NCCL timeouts is the SPMD programming model as implemented in PyTorch. If different ranks accidentally issue collectives such as \texttt{All-Reduce} in the wrong order, the job will deadlock, resulting in a NCCL timeout. Debugging a NCCL timeout, therefore, first starts with determining if the training script was buggy, adding a confounding factor to tracking down infrastructure instability. Dynamically detecting incorrect programs and raising exceptions rather than deadlocking would improve stability. 
    Alternatively, one can aim to eliminate the possibility of mismatched collectives entirely. For instance, Pathways~\cite{barham2022pathwaysasynchronousdistributeddataflow}, introduces a single point where communication is scheduled ensuring each machine schedules communication consistently.
    On the algorithmic side, there is renewed interest in algorithms that are asynchronous.
    Past systems embraced asynchronous gradient exchange as a performance optimization~\cite{distbelief,project_adam,hogwild}.
    Current research trends look at this as a potentially cross-cluster federated learning problem~\cite{dilico}.
     
\section{Related Work}%
\label{sec:related_work}%
\vspace{-0.1cm}

\textbf{ML Infrastructure.} A growing interest in ML has led to work investigating infrastructure failures as well as scheduler effects, specifically for ML clusters~\cite{jeon2019,ai_enabling_workloads}. This includes analyzing aspects from job sizes to failure properties, predominantly at maximum job scales reaching into the tens of GPUs.
IBM's AI infrastructure and failure taxonomy was introduced in \cite{gershon2024infrastructurepoweringibmsgen}---we provide a similarly motivated taxonomy and quantify the failures in detail.
The design and operation of Meta's production scheduler, MAST, was also recently studied~\cite{MAST}, demonstrating a need for datacenter-level load balancing.
The rise of LLMs has recently spurred a number of recent works specifically focusing on reliability for ML training~\cite{llama3-model-card, gemini, megascale}, assuming orders of magnitude larger scale than previously seen in general-purpose ML experiments.
Recent work has analyzed LLM-specific datacenter workloads~\cite{llm_datacenter_2024}, finding that evaluation jobs are important.
In comparison, our work serves the entire range of job sizes from 1 to 4k GPUs and does so with a variety of cutting-edge machine learning research workloads---emphasizing general purpose reliability techniques.

\textbf{Model-System Codesign.} LLM-specific parallelization techniques, such as Megatron-LM~\cite{shoeybi2020megatronlmtrainingmultibillionparameter}, have been developed---naturally producing a hierarchical mapping of data and model parallelism to common network topologies~\cite{wang2023optimized}. LLM failures and mitigations at ByteDance were studied in Megascale~\cite{megascale}. A similar study was performed for Alibaba training systems in Unicron~\cite{he2023unicroneconomizingselfhealingllm}, and a specialized HPN network was proposed to mitigate networking-related failures~\cite{alibaba_hpn}. 
LLaMa 3 also demonstrated benefits from software-hardware codesign to improve resilience~\cite{llama3-model-card}. 
In addition, fault-tolerant, resilient training has also been explored for other large deep learning model development~\cite{maeng-mlsys2020,zhang-vldb2023}. 
While our work tackles the same model training resilience as well as general purpose cluster reliability, our experience is unique in that we operate at both ends of the scale, motivating solutions that are practical to be deployed in a black-box manner, yet still battle tested at the scale of four thousand GPUs.
Finally, our workloads are diverse~\cite{touvron2023llamaopenefficientfoundation,audiobox,seg,emu,nllb, codegen}, spanning vision, language, and mixed-modality models in a rapidly evolving research setting. %
\section{Conclusion}
\vspace{-0.1cm}
We share our experience in operating two large-scale ML research clusters. Our analysis indicates that state-of-the-art research clusters operate at orders of magnitude larger scale than previously demonstrated, motivating an increased emphasis on large-scale performance and reliability.
This paper takes a data-driven approach to quantify and mitigate the impact of failures, ranging from hardware, to system software, to framework-level resiliency improvements. \arxivonly{ 
\section*{Acknowledgement}
\vspace{-0.1cm}
We thank our industry partners, especially Nvidia, for pioneering debugging methodologies when we encountered a fresh set of challenges.
Close collaborations enabled us to debug failures as well as develop new health-checks to improve the cluster reliability over time.
We especially thank Nvidia's Tel Aviv/Yokne'am Illit teams
for helping us triage failures and bandwidth performance issues at scale and for delivering features that enabled better cluster management. We thank
Devendra Ayalasomayajula,
Kevin Schlichter,
Nandini Shankarappa,
Sam Simcoe,
Dan Waxman,
and our colleagues at Meta for their work in enabling our infrastructure and making this work possible.
Lastly, we thank Kim Hazelwood and Leo Huang for their support and guidance.
 }

\bibliographystyle{IEEEtranS}
\bibliography{refs}

% Generated by IEEEtranS.bst, version: 1.13 (2008/09/30)
\begin{thebibliography}{10}
\providecommand{\url}[1]{#1}
\csname url@samestyle\endcsname
\providecommand{\newblock}{\relax}
\providecommand{\bibinfo}[2]{#2}
\providecommand{\BIBentrySTDinterwordspacing}{\spaceskip=0pt\relax}
\providecommand{\BIBentryALTinterwordstretchfactor}{4}
\providecommand{\BIBentryALTinterwordspacing}{\spaceskip=\fontdimen2\font plus
\BIBentryALTinterwordstretchfactor\fontdimen3\font minus
  \fontdimen4\font\relax}
\providecommand{\BIBforeignlanguage}[2]{{%
\expandafter\ifx\csname l@#1\endcsname\relax
\typeout{** WARNING: IEEEtranS.bst: No hyphenation pattern has been}%
\typeout{** loaded for the language `#1'. Using the pattern for}%
\typeout{** the default language instead.}%
\else
\language=\csname l@#1\endcsname
\fi
#2}}
\providecommand{\BIBdecl}{\relax}
\BIBdecl

\bibitem{meta_genai_cluster}
``Building meta’s genai infrastructure,''
  \url{https://engineering.fb.com/2024/03/12/data-center-engineering/building-metas-genai-infrastructure/},
  (Accessed on 8/5/2024).

\bibitem{tpuv5}
``Enabling next-generation ai workloads: Announcing tpu v5p and ai
  hypercomputer,''
  \url{https://cloud.google.com/blog/products/ai-machine-learning/introducing-cloud-tpu-v5p-and-ai-hypercomputer},
  (Accessed on 7/17/2024).

\bibitem{ml_productivity_goodput}
``Introducing ml productivity goodput: a metric to measure ai system
  efficiency,''
  \url{https://cloud.google.com/blog/products/ai-machine-learning/goodput-metric-as-measure-of-ml-productivity},
  (Accessed on 9/9/2024).

\bibitem{ai_rsc}
``Introducing the ai research supercluster — meta’s cutting-edge ai
  supercomputer for ai research,'' \url{https://ai.meta.com/blog/ai-rsc/},
  (Accessed on 7/18/2024).

\bibitem{SLURMPriority}
``Multifactor priority plugin,''
  \url{https://slurm.schedmd.com/priority_multifactor.html#general}, (Accessed
  on 8/7/2024).

\bibitem{nccl_tests}
``Nccl-tests,'' \url{https://github.com/NVIDIA/nccl-tests}, (Accessed on
  8/06/2024).

\bibitem{DGXA100Datasheet}
``Nvidia dgx a100,''
  \url{https://images.nvidia.com/aem-dam/Solutions/Data-Center/nvidia-dgx-a100-datasheet.pdf},
  (Accessed on 8/7/2024).

\bibitem{GB200}
``Nvidia gb200 nvl72,''
  \url{https://www.nvidia.com/en-us/data-center/gb200-nvl72/}, (Accessed on
  9/13/2024).

\bibitem{xid}
``Nvidia xid error messages,''
  \url{https://docs.nvidia.com/deploy/pdf/XID_Errors.pdf}, (Accessed on
  7/19/2024).

\bibitem{SLURMPrologEpilog}
``Prolog and epilog guide,''
  \url{https://slurm.schedmd.com/prolog_epilog.html}, (Accessed on 8/7/2024).

\bibitem{flightrecorder}
``(prototype) flight recorder for debugging stuck jobs,''
  \url{https://pytorch.org/tutorials/prototype/flight_recorder_tutorial.html},
  (Accessed on 11/4/2024).

\bibitem{shield}
``The shield: Self-healing interconnect,''
  \url{https://network.nvidia.com/related-docs/whitepapers/WP_Mellanox_SHIELD.pdf},
  (Accessed on 7/18/2024).

\bibitem{submitit}
``Submit it!'' \url{https://github.com/facebookincubator/submitit}, (Accessed
  on 9/15/2024).

\bibitem{pytorch2}
J.~Ansel, E.~Yang, H.~He, N.~Gimelshein, A.~Jain, M.~Voznesensky, B.~Bao,
  P.~Bell, D.~Berard, E.~Burovski, G.~Chauhan, A.~Chourdia, W.~Constable,
  A.~Desmaison, Z.~DeVito, E.~Ellison, W.~Feng, J.~Gong, M.~Gschwind, B.~Hirsh,
  S.~Huang, K.~Kalambarkar, L.~Kirsch, M.~Lazos, M.~Lezcano, Y.~Liang,
  J.~Liang, Y.~Lu, C.~K. Luk, B.~Maher, Y.~Pan, C.~Puhrsch, M.~Reso,
  M.~Saroufim, M.~Y. Siraichi, H.~Suk, S.~Zhang, M.~Suo, P.~Tillet, X.~Zhao,
  E.~Wang, K.~Zhou, R.~Zou, X.~Wang, A.~Mathews, W.~Wen, G.~Chanan, P.~Wu, and
  S.~Chintala, ``Pytorch 2: Faster machine learning through dynamic python
  bytecode transformation and graph compilation,'' in \emph{Proceedings of the
  29th ACM International Conference on Architectural Support for Programming
  Languages and Operating Systems, Volume 2}, 2024.

\bibitem{barham2022pathwaysasynchronousdistributeddataflow}
P.~Barham, A.~Chowdhery, J.~Dean, S.~Ghemawat, S.~Hand, D.~Hurt, M.~Isard,
  H.~Lim, R.~Pang, S.~Roy, B.~Saeta, P.~Schuh, R.~Sepassi, L.~E. Shafey, C.~A.
  Thekkath, and Y.~Wu, ``Pathways: Asynchronous distributed dataflow for ml,''
  in \emph{Proceedings of Machine Learning and Systems}, 2022.

\bibitem{Bautista-Gomez_etal_2024}
\BIBentryALTinterwordspacing
L.~Bautista-Gomez, A.~Benoit, S.~Di, T.~Herault, Y.~Robert, and H.~Sun, ``A
  survey on checkpointing strategies: Should we always checkpoint à la
  young/daly?'' \emph{Future Generation Computer Systems}, vol. 161, pp.
  315--328, 2024. [Online]. Available:
  \url{https://www.sciencedirect.com/science/article/pii/S0167739X24003777}
\BIBentrySTDinterwordspacing

\bibitem{Benoit_etal_2022}
\BIBentryALTinterwordspacing
A.~Benoit, Y.~Du, T.~Herault, L.~Marchal, G.~Pallez, L.~Perotin, Y.~Robert,
  H.~Sun, and F.~Vivien, ``Checkpointing \`{a} la young/daly: An overview,'' in
  \emph{Proceedings of the 2022 Fourteenth International Conference on
  Contemporary Computing}, ser. IC3-2022.\hskip 1em plus 0.5em minus
  0.4em\relax New York, NY, USA: Association for Computing Machinery, 2022, p.
  701–710. [Online]. Available: \url{https://doi.org/10.1145/3549206.3549328}
\BIBentrySTDinterwordspacing

\bibitem{SDC-google}
R.~Bonderson, ``Training in turmoil: Silent data corruption in systems at
  scale,'' 2021, for submission to an invited talk at the International Test
  Conference, in a "Silicon Lifecycle Management" workshop. Conference is
  October 10-15, with this presentation / talk sometime on the 15th.

\bibitem{project_adam}
\BIBentryALTinterwordspacing
T.~Chilimbi, Y.~Suzue, J.~Apacible, and K.~Kalyanaraman, ``Project adam:
  Building an efficient and scalable deep learning training system,'' in
  \emph{11th USENIX Symposium on Operating Systems Design and Implementation
  (OSDI 14)}.\hskip 1em plus 0.5em minus 0.4em\relax Broomfield, CO: USENIX
  Association, Oct. 2014, pp. 571--582. [Online]. Available:
  \url{https://www.usenix.org/conference/osdi14/technical-sessions/presentation/chilimbi}
\BIBentrySTDinterwordspacing

\bibitem{MAST}
\BIBentryALTinterwordspacing
A.~Choudhury, Y.~Wang, T.~Pelkonen, K.~Srinivasan, A.~Jain, S.~Lin, D.~David,
  S.~Soleimanifard, M.~Chen, A.~Yadav, R.~Tijoriwala, D.~Samoylov, and C.~Tang,
  ``{MAST}: Global scheduling of {ML} training across {Geo-Distributed}
  datacenters at hyperscale,'' in \emph{18th USENIX Symposium on Operating
  Systems Design and Implementation (OSDI 24)}.\hskip 1em plus 0.5em minus
  0.4em\relax Santa Clara, CA: USENIX Association, Jul. 2024, pp. 563--580.
  [Online]. Available:
  \url{https://www.usenix.org/conference/osdi24/presentation/choudhury}
\BIBentrySTDinterwordspacing

\bibitem{palm}
\BIBentryALTinterwordspacing
A.~Chowdhery, S.~Narang, J.~Devlin, M.~Bosma, G.~Mishra, A.~Roberts, P.~Barham,
  H.~W. Chung, C.~Sutton, S.~Gehrmann, P.~Schuh, K.~Shi, S.~Tsvyashchenko,
  J.~Maynez, A.~Rao, P.~Barnes, Y.~Tay, N.~Shazeer, V.~Prabhakaran, E.~Reif,
  N.~Du, B.~Hutchinson, R.~Pope, J.~Bradbury, J.~Austin, M.~Isard, G.~Gur-Ari,
  P.~Yin, T.~Duke, A.~Levskaya, S.~Ghemawat, S.~Dev, H.~Michalewski, X.~Garcia,
  V.~Misra, K.~Robinson, L.~Fedus, D.~Zhou, D.~Ippolito, D.~Luan, H.~Lim,
  B.~Zoph, A.~Spiridonov, R.~Sepassi, D.~Dohan, S.~Agrawal, M.~Omernick, A.~M.
  Dai, T.~S. Pillai, M.~Pellat, A.~Lewkowycz, E.~Moreira, R.~Child, O.~Polozov,
  K.~Lee, Z.~Zhou, X.~Wang, B.~Saeta, M.~Diaz, O.~Firat, M.~Catasta, J.~Wei,
  K.~Meier-Hellstern, D.~Eck, J.~Dean, S.~Petrov, and N.~Fiedel, ``Palm:
  Scaling language modeling with pathways,'' \emph{Journal of Machine Learning
  Research}, vol.~24, no. 240, pp. 1--113, 2023. [Online]. Available:
  \url{http://jmlr.org/papers/v24/22-1144.html}
\BIBentrySTDinterwordspacing

\bibitem{emu}
\BIBentryALTinterwordspacing
X.~Dai, J.~Hou, C.-Y. Ma, S.~Tsai, J.~Wang, R.~Wang, P.~Zhang, S.~Vandenhende,
  X.~Wang, A.~Dubey, M.~Yu, A.~Kadian, F.~Radenovic, D.~Mahajan, K.~Li,
  Y.~Zhao, V.~Petrovic, M.~K. Singh, S.~Motwani, Y.~Wen, Y.~Song, R.~Sumbaly,
  V.~Ramanathan, Z.~He, P.~Vajda, and D.~Parikh, ``{Emu: Enhancing Image
  Generation Models Using Photogenic Needles in a Haystack},'' 2023. [Online].
  Available: \url{https://arxiv.org/abs/2309.15807}
\BIBentrySTDinterwordspacing

\bibitem{Daly_2006}
\BIBentryALTinterwordspacing
J.~Daly, ``A higher order estimate of the optimum checkpoint interval for
  restart dumps,'' \emph{Future Generation Computer Systems}, vol.~22, no.~3,
  pp. 303--312, 2006. [Online]. Available:
  \url{https://www.sciencedirect.com/science/article/pii/S0167739X04002213}
\BIBentrySTDinterwordspacing

\bibitem{distbelief}
\BIBentryALTinterwordspacing
J.~Dean, G.~Corrado, R.~Monga, K.~Chen, M.~Devin, M.~Mao, M.~a. Ranzato,
  A.~Senior, P.~Tucker, K.~Yang, Q.~Le, and A.~Ng, ``Large scale distributed
  deep networks,'' in \emph{Advances in Neural Information Processing Systems},
  vol.~25, 2012. [Online]. Available:
  \url{https://proceedings.neurips.cc/paper_files/paper/2012/file/6aca97005c68f1206823815f66102863-Paper.pdf}
\BIBentrySTDinterwordspacing

\bibitem{SDC-meta}
\BIBentryALTinterwordspacing
H.~D. Dixit, S.~Pendharkar, M.~Beadon, C.~Mason, T.~Chakravarthy, B.~Muthiah,
  and S.~Sankar, ``Silent data corruptions at scale,'' 2021. [Online].
  Available: \url{https://arxiv.org/abs/2102.11245}
\BIBentrySTDinterwordspacing

\bibitem{dilico}
\BIBentryALTinterwordspacing
A.~Douillard, Q.~Feng, A.~A. Rusu, R.~Chhaparia, Y.~Donchev, A.~Kuncoro,
  M.~Ranzato, A.~Szlam, and J.~Shen, ``Diloco: Distributed low-communication
  training of language models,'' in \emph{2nd Workshop on Advancing Neural
  Network Training: Computational Efficiency, Scalability, and Resource
  Optimization (WANT@ICML 2024)}, 2024. [Online]. Available:
  \url{https://openreview.net/forum?id=pICSfWkJIk}
\BIBentrySTDinterwordspacing

\bibitem{epoch_hardware_failures_2024}
\BIBentryALTinterwordspacing
A.~Erben and E.~Erdil, ``Hardware failures won’t limit ai scaling,'' 2024,
  accessed: 2024-12-06. [Online]. Available:
  \url{https://epoch.ai/blog/hardware-failures-wont-limit-ai-scaling}
\BIBentrySTDinterwordspacing

\bibitem{gemini}
\BIBentryALTinterwordspacing
{Gemini Team et. al.}, ``Gemini: A family of highly capable multimodal
  models,'' 2024. [Online]. Available: \url{https://arxiv.org/abs/2312.11805}
\BIBentrySTDinterwordspacing

\bibitem{gershon2024infrastructurepoweringibmsgen}
\BIBentryALTinterwordspacing
T.~Gershon, S.~Seelam, B.~Belgodere, M.~Bonilla, L.~Hoang, D.~Barnett, I.-H.
  Chung, A.~Mohan, M.-H. Chen, L.~Luo, R.~Walkup, C.~Evangelinos, S.~Salaria,
  M.~Dombrowa, Y.~Park, A.~Kayi, L.~Schour, A.~Alim, A.~Sydney, P.~Maniotis,
  L.~Schares, B.~Metzler, B.~Karacali-Akyamac, S.~Wen, T.~Chiba,
  S.~Choochotkaew, T.~Yoshimura, C.~Misale, T.~Elengikal, K.~O. Connor, Z.~Liu,
  R.~Molina, L.~Schneidenbach, J.~Caden, C.~Laibinis, C.~Fonseca, V.~Tarasov,
  S.~Sundararaman, F.~Schmuck, S.~Guthridge, J.~Cohn, M.~Eshel, P.~Muench,
  R.~Liu, W.~Pointer, D.~Wyskida, B.~Krull, R.~Rose, B.~Wolfe, W.~Cornejo,
  J.~Walter, C.~Malone, C.~Perucci, F.~Franco, N.~Hinds, B.~Calio, P.~Druyan,
  R.~Kilduff, J.~Kienle, C.~McStay, A.~Figueroa, M.~Connolly, E.~Fost, G.~Roma,
  J.~Fonseca, I.~Levy, M.~Payne, R.~Schenkel, A.~Malki, L.~Schneider,
  A.~Narkhede, S.~Moshref, A.~Kisin, O.~Dodin, B.~Rippon, H.~Wrieth, J.~Ganci,
  J.~Colino, D.~Habeger-Rose, R.~Pandey, A.~Gidh, A.~Gaur, D.~Patterson,
  S.~Salmani, R.~Varma, R.~Rumana, S.~Sharma, A.~Gaur, M.~Mishra, R.~Panda,
  A.~Prasad, M.~Stallone, G.~Zhang, Y.~Shen, D.~Cox, R.~Puri, D.~Agrawal,
  D.~Thorstensen, J.~Belog, B.~Tang, S.~K. Gupta, A.~Biswas, A.~Maheshwari,
  E.~Gampel, J.~V. Patten, M.~Runion, S.~Kaki, Y.~Bogin, B.~Reitz, S.~Pritko,
  S.~Najam, S.~Nambala, R.~Chirra, R.~Welp, F.~DiMitri, F.~Telles, A.~Arvelo,
  K.~Chu, E.~Seminaro, A.~Schram, F.~Eickhoff, W.~Hanson, E.~Mckeever,
  D.~Joseph, P.~Chaudhary, P.~Shivam, P.~Chaudhary, W.~Jones, R.~Guthrie,
  C.~Bostic, R.~Islam, S.~Duersch, W.~Sawdon, J.~Lewars, M.~Klos, M.~Spriggs,
  B.~McMillan, G.~Gao, A.~Kamra, G.~Singh, M.~Curry, T.~Katarki, J.~Talerico,
  Z.~Shi, S.~S. Malleni, and E.~Gallen, ``The infrastructure powering ibm's gen
  ai model development,'' 2024. [Online]. Available:
  \url{https://arxiv.org/abs/2407.05467}
\BIBentrySTDinterwordspacing

\bibitem{HANSEN_2003}
F.~Hansen and G.~K. Pedersen, ``Jensen’s operator inequality,''
  \emph{Bulletin of the London Mathematical Society}, vol.~35, no.~04, p.
  553–564, 2003.

\bibitem{Harchol-Balter-2002}
\BIBentryALTinterwordspacing
M.~Harchol-Balter, K.~Sigman, and A.~Wierman, ``Asymptotic convergence of
  scheduling policies with respect to slowdown,'' \emph{Performance
  Evaluation}, vol.~49, no.~1, pp. 241--256, 2002, performance 2002. [Online].
  Available:
  \url{https://www.sciencedirect.com/science/article/pii/S0166531602001323}
\BIBentrySTDinterwordspacing

\bibitem{he2023unicroneconomizingselfhealingllm}
\BIBentryALTinterwordspacing
T.~He, X.~Li, Z.~Wang, K.~Qian, J.~Xu, W.~Yu, and J.~Zhou, ``Unicron:
  Economizing self-healing llm training at scale,'' 2023. [Online]. Available:
  \url{https://arxiv.org/abs/2401.00134}
\BIBentrySTDinterwordspacing

\bibitem{understanding_hwfailure_sdc}
\BIBentryALTinterwordspacing
Y.~He, M.~Hutton, S.~Chan, R.~De~Gruijl, R.~Govindaraju, N.~Patil, and Y.~Li,
  ``Understanding and mitigating hardware failures in deep learning training
  systems,'' in \emph{Proceedings of the 50th Annual International Symposium on
  Computer Architecture}, ser. ISCA '23.\hskip 1em plus 0.5em minus 0.4em\relax
  New York, NY, USA: Association for Computing Machinery, 2023. [Online].
  Available: \url{https://doi.org/10.1145/3579371.3589105}
\BIBentrySTDinterwordspacing

\bibitem{hsia-isca2024}
S.~Hsia, A.~Golden, B.~Acun, N.~Ardalani, Z.~DeVito, G.-Y. Wei, D.~Brooks, and
  C.-J. Wu, ``Mad-max beyond single-node: Enabling large machine learning model
  acceleration on distributed systems,'' in \emph{2024 ACM/IEEE 51st Annual
  International Symposium on Computer Architecture (ISCA)}, 2024, pp. 818--833.

\bibitem{llm_datacenter_2024}
\BIBentryALTinterwordspacing
Q.~Hu, Z.~Ye, Z.~Wang, G.~Wang, M.~Zhang, Q.~Chen, P.~Sun, D.~Lin, X.~Wang,
  Y.~Luo, Y.~Wen, and T.~Zhang, ``Characterization of large language model
  development in the datacenter,'' in \emph{21st USENIX Symposium on Networked
  Systems Design and Implementation (NSDI 24)}.\hskip 1em plus 0.5em minus
  0.4em\relax Santa Clara, CA: USENIX Association, Apr. 2024, pp. 709--729.
  [Online]. Available:
  \url{https://www.usenix.org/conference/nsdi24/presentation/hu}
\BIBentrySTDinterwordspacing

\bibitem{ipmi}
Intel, Hewlett-Packard, NEC, and Dell, ``Intelligent platform management
  interface specification second generation v2.0,'' Intel, Tech. Rep., 2013.

\bibitem{jeon2019}
\BIBentryALTinterwordspacing
M.~Jeon, S.~Venkataraman, A.~Phanishayee, J.~Qian, W.~Xiao, and F.~Yang,
  ``Analysis of {Large-Scale} {Multi-Tenant} {GPU} clusters for {DNN} training
  workloads,'' in \emph{2019 USENIX Annual Technical Conference (USENIX ATC
  19)}.\hskip 1em plus 0.5em minus 0.4em\relax Renton, WA: USENIX Association,
  Jul. 2019, pp. 947--960. [Online]. Available:
  \url{https://www.usenix.org/conference/atc19/presentation/jeon}
\BIBentrySTDinterwordspacing

\bibitem{megascale}
\BIBentryALTinterwordspacing
Z.~Jiang, H.~Lin, Y.~Zhong, Q.~Huang, Y.~Chen, Z.~Zhang, Y.~Peng, X.~Li,
  C.~Xie, S.~Nong, Y.~Jia, S.~He, H.~Chen, Z.~Bai, Q.~Hou, S.~Yan, D.~Zhou,
  Y.~Sheng, Z.~Jiang, H.~Xu, H.~Wei, Z.~Zhang, P.~Nie, L.~Zou, S.~Zhao,
  L.~Xiang, Z.~Liu, Z.~Li, X.~Jia, J.~Ye, X.~Jin, and X.~Liu, ``{MegaScale}:
  Scaling large language model training to more than 10,000 {GPUs},'' in
  \emph{21st USENIX Symposium on Networked Systems Design and Implementation
  (NSDI 24)}.\hskip 1em plus 0.5em minus 0.4em\relax Santa Clara, CA: USENIX
  Association, Apr. 2024, pp. 745--760. [Online]. Available:
  \url{https://www.usenix.org/conference/nsdi24/presentation/jiang-ziheng}
\BIBentrySTDinterwordspacing

\bibitem{tpuv4}
N.~Jouppi, G.~Kurian, S.~Li, P.~Ma, R.~Nagarajan, L.~Nai, N.~Patil,
  S.~Subramanian, A.~Swing, B.~Towles, C.~Young, X.~Zhou, Z.~Zhou, and D.~A.
  Patterson, ``Tpu v4: An optically reconfigurable supercomputer for machine
  learning with hardware support for embeddings,'' in \emph{Proceedings of the
  50th Annual International Symposium on Computer Architecture}, 2023.

\bibitem{tpuv1}
\BIBentryALTinterwordspacing
N.~P. Jouppi, C.~Young, N.~Patil, D.~Patterson, G.~Agrawal, R.~Bajwa, S.~Bates,
  S.~Bhatia, N.~Boden, A.~Borchers, R.~Boyle, P.-l. Cantin, C.~Chao, C.~Clark,
  J.~Coriell, M.~Daley, M.~Dau, J.~Dean, B.~Gelb, T.~V. Ghaemmaghami,
  R.~Gottipati, W.~Gulland, R.~Hagmann, C.~R. Ho, D.~Hogberg, J.~Hu, R.~Hundt,
  D.~Hurt, J.~Ibarz, A.~Jaffey, A.~Jaworski, A.~Kaplan, H.~Khaitan,
  D.~Killebrew, A.~Koch, N.~Kumar, S.~Lacy, J.~Laudon, J.~Law, D.~Le, C.~Leary,
  Z.~Liu, K.~Lucke, A.~Lundin, G.~MacKean, A.~Maggiore, M.~Mahony, K.~Miller,
  R.~Nagarajan, R.~Narayanaswami, R.~Ni, K.~Nix, T.~Norrie, M.~Omernick,
  N.~Penukonda, A.~Phelps, J.~Ross, M.~Ross, A.~Salek, E.~Samadiani, C.~Severn,
  G.~Sizikov, M.~Snelham, J.~Souter, D.~Steinberg, A.~Swing, M.~Tan,
  G.~Thorson, B.~Tian, H.~Toma, E.~Tuttle, V.~Vasudevan, R.~Walter, W.~Wang,
  E.~Wilcox, and D.~H. Yoon, ``In-datacenter performance analysis of a tensor
  processing unit,'' in \emph{Proceedings of the 44th Annual International
  Symposium on Computer Architecture}, ser. ISCA '17.\hskip 1em plus 0.5em
  minus 0.4em\relax New York, NY, USA: Association for Computing Machinery,
  2017, p. 1–12. [Online]. Available:
  \url{https://doi.org/10.1145/3079856.3080246}
\BIBentrySTDinterwordspacing

\bibitem{seg}
\BIBentryALTinterwordspacing
A.~Kirillov, E.~Mintun, N.~Ravi, H.~Mao, C.~Rolland, L.~Gustafson, T.~Xiao,
  S.~Whitehead, A.~C. Berg, W.-Y. Lo, P.~Dollár, and R.~Girshick, ``{Segment
  Anything},'' 2023. [Online]. Available:
  \url{https://arxiv.org/abs/2304.02643}
\BIBentrySTDinterwordspacing

\bibitem{korthikanti2022reducingactivationrecomputationlarge}
\BIBentryALTinterwordspacing
V.~Korthikanti, J.~Casper, S.~Lym, L.~McAfee, M.~Andersch, M.~Shoeybi, and
  B.~Catanzaro, ``Reducing activation recomputation in large transformer
  models,'' 2022. [Online]. Available: \url{https://arxiv.org/abs/2205.05198}
\BIBentrySTDinterwordspacing

\bibitem{cielo_memory}
S.~Levy, K.~B. Ferreira, N.~DeBardeleben, T.~Siddiqua, V.~Sridharan, and
  E.~Baseman, ``Lessons learned from memory errors observed over the lifetime
  of cielo,'' in \emph{SC18: International Conference for High Performance
  Computing, Networking, Storage and Analysis}, 2018, pp. 554--565.

\bibitem{ai_enabling_workloads}
B.~Li, R.~Arora, S.~Samsi, T.~Patel, W.~Arcand, D.~Bestor, C.~Byun, R.~B. Roy,
  B.~Bergeron, J.~Holodnak, M.~Houle, M.~Hubbell, M.~Jones, J.~Kepner,
  A.~Klein, P.~Michaleas, J.~McDonald, L.~Milechin, J.~Mullen, A.~Prout,
  B.~Price, A.~Reuther, A.~Rosa, M.~Weiss, C.~Yee, D.~Edelman, A.~Vanterpool,
  A.~Cheng, V.~Gadepally, and D.~Tiwari, ``{AI-Enabling Workloads on
  Large-Scale GPU-Accelerated System: Characterization, Opportunities, and
  Implications},'' in \emph{2022 IEEE International Symposium on
  High-Performance Computer Architecture (HPCA)}, 2022, pp. 1224--1237.

\bibitem{drdna}
\BIBentryALTinterwordspacing
D.~Ma, F.~Lin, A.~Desmaison, J.~Coburn, D.~Moore, S.~Sankar, and X.~Jiao, ``Dr.
  dna: Combating silent data corruptions in deep learning using distribution of
  neuron activations,'' in \emph{Proceedings of the 29th ACM International
  Conference on Architectural Support for Programming Languages and Operating
  Systems, Volume 3}, ser. ASPLOS '24.\hskip 1em plus 0.5em minus 0.4em\relax
  New York, NY, USA: Association for Computing Machinery, 2024, p. 239–252.
  [Online]. Available: \url{https://doi.org/10.1145/3620666.3651349}
\BIBentrySTDinterwordspacing

\bibitem{maeng-mlsys2020}
K.~Maeng, S.~Bharuka, I.~Gao, M.~C. Jeffrey, V.~Saraph, B.-Y. Su, C.~Trippel,
  J.~Yang, M.~Rabbat, B.~Lucia, and C.-J. Wu, ``Cpr: Understanding and
  improving failure tolerant training for deep learning recommendation with
  partial recovery,'' in \emph{Proceedings of Machine Learning and Systems},
  2021.

\bibitem{llama3-model-card}
\BIBentryALTinterwordspacing
Meta, ``The official {Meta Llama 3 GitHub site},'' 2024. [Online]. Available:
  \url{https://github.com/meta-llama/llama3}
\BIBentrySTDinterwordspacing

\bibitem{nllb}
\BIBentryALTinterwordspacing
{NLLB Team}, M.~R. Costa-jussà, J.~Cross, O.~Çelebi, M.~Elbayad, K.~Heafield,
  K.~Heffernan, E.~Kalbassi, J.~Lam, D.~Licht, J.~Maillard, A.~Sun, S.~Wang,
  G.~Wenzek, A.~Youngblood, B.~Akula, L.~Barrault, G.~M. Gonzalez, P.~Hansanti,
  J.~Hoffman, S.~Jarrett, K.~R. Sadagopan, D.~Rowe, S.~Spruit, C.~Tran,
  P.~Andrews, N.~F. Ayan, S.~Bhosale, S.~Edunov, A.~Fan, C.~Gao, V.~Goswami,
  F.~Guzmán, P.~Koehn, A.~Mourachko, C.~Ropers, S.~Saleem, H.~Schwenk, and
  J.~Wang, ``{No Language Left Behind: Scaling Human-Centered Machine
  Translation},'' 2022. [Online]. Available:
  \url{https://arxiv.org/abs/2207.04672}
\BIBentrySTDinterwordspacing

\bibitem{gpt4}
\BIBentryALTinterwordspacing
{OpenAI et. al.}, ``Gpt-4 technical report,'' 2024. [Online]. Available:
  \url{https://arxiv.org/abs/2303.08774}
\BIBentrySTDinterwordspacing

\bibitem{alibaba_hpn}
K.~Qian, Y.~Xi, J.~Caoa, J.~Gao, Y.~Xu, Y.~Guan, B.~Fu, X.~Shi, F.~Zhu,
  R.~Miao, C.~Wang, P.~Wang, P.~Zhang, X.~Zeng, Z.~Yao, E.~Zhai, and D.~Cai,
  ``Alibaba hpn: A data center network for large language model training,'' in
  \emph{SIGCOMM}, 2024.

\bibitem{hogwild}
\BIBentryALTinterwordspacing
B.~Recht, C.~Re, S.~Wright, and F.~Niu, ``Hogwild!: A lock-free approach to
  parallelizing stochastic gradient descent,'' in \emph{Advances in Neural
  Information Processing Systems}, vol.~24, 2011. [Online]. Available:
  \url{https://proceedings.neurips.cc/paper_files/paper/2011/file/218a0aefd1d1a4be65601cc6ddc1520e-Paper.pdf}
\BIBentrySTDinterwordspacing

\bibitem{google_trace_2012}
\BIBentryALTinterwordspacing
C.~Reiss, A.~Tumanov, G.~R. Ganger, R.~H. Katz, and M.~A. Kozuch,
  ``Heterogeneity and dynamicity of clouds at scale: Google trace analysis,''
  in \emph{Proceedings of the Third ACM Symposium on Cloud Computing}, ser.
  SoCC '12.\hskip 1em plus 0.5em minus 0.4em\relax New York, NY, USA:
  Association for Computing Machinery, 2012. [Online]. Available:
  \url{https://doi.org/10.1145/2391229.2391236}
\BIBentrySTDinterwordspacing

\bibitem{codegen}
B.~Roziere, M.-A. Lachaux, L.~Chanussot, and G.~Lample, ``{Unsupervised
  translation of programming languages},'' \emph{Advances in Neural Information
  Processing Systems}, vol.~33, 2020.

\bibitem{shoeybi2020megatronlmtrainingmultibillionparameter}
\BIBentryALTinterwordspacing
M.~Shoeybi, M.~Patwary, R.~Puri, P.~LeGresley, J.~Casper, and B.~Catanzaro,
  ``Megatron-lm: Training multi-billion parameter language models using model
  parallelism,'' 2020. [Online]. Available:
  \url{https://arxiv.org/abs/1909.08053}
\BIBentrySTDinterwordspacing

\bibitem{gpu_failures}
D.~Tiwari, S.~Gupta, J.~Rogers, D.~Maxwell, P.~Rech, S.~Vazhkudai, D.~Oliveira,
  D.~Londo, N.~DeBardeleben, P.~Navaux, L.~Carro, and A.~Bland,
  ``{Understanding GPU errors on large-scale HPC systems and the implications
  for system design and operation},'' in \emph{2015 IEEE 21st International
  Symposium on High Performance Computer Architecture (HPCA)}, 2015, pp.
  331--342.

\bibitem{touvron2023llamaopenefficientfoundation}
\BIBentryALTinterwordspacing
H.~Touvron, T.~Lavril, G.~Izacard, X.~Martinet, M.-A. Lachaux, T.~Lacroix,
  B.~Rozière, N.~Goyal, E.~Hambro, F.~Azhar, A.~Rodriguez, A.~Joulin,
  E.~Grave, and G.~Lample, ``Llama: Open and efficient foundation language
  models,'' 2023. [Online]. Available: \url{https://arxiv.org/abs/2302.13971}
\BIBentrySTDinterwordspacing

\bibitem{Valiant_BSP}
\BIBentryALTinterwordspacing
L.~G. Valiant, ``A bridging model for parallel computation,'' \emph{Commun.
  ACM}, vol.~33, no.~8, p. 103–111, Aug. 1990. [Online]. Available:
  \url{https://doi.org/10.1145/79173.79181}
\BIBentrySTDinterwordspacing

\bibitem{borg}
\BIBentryALTinterwordspacing
A.~Verma, L.~Pedrosa, M.~Korupolu, D.~Oppenheimer, E.~Tune, and J.~Wilkes,
  ``Large-scale cluster management at google with borg,'' in \emph{Proceedings
  of the Tenth European Conference on Computer Systems}, ser. EuroSys
  '15.\hskip 1em plus 0.5em minus 0.4em\relax New York, NY, USA: Association
  for Computing Machinery, 2015. [Online]. Available:
  \url{https://doi.org/10.1145/2741948.2741964}
\BIBentrySTDinterwordspacing

\bibitem{audiobox}
\BIBentryALTinterwordspacing
A.~Vyas, B.~Shi, M.~Le, A.~Tjandra, Y.-C. Wu, B.~Guo, J.~Zhang, X.~Zhang,
  R.~Adkins, W.~Ngan, J.~Wang, I.~Cruz, B.~Akula, A.~Akinyemi, B.~Ellis,
  R.~Moritz, Y.~Yungster, A.~Rakotoarison, L.~Tan, C.~Summers, C.~Wood,
  J.~Lane, M.~Williamson, and W.-N. Hsu, ``{Audiobox: Unified Audio Generation
  with Natural Language Prompts},'' 2023. [Online]. Available:
  \url{https://arxiv.org/abs/2312.15821}
\BIBentrySTDinterwordspacing

\bibitem{wang2023optimized}
\BIBentryALTinterwordspacing
W.~Wang, M.~Ghobadi, K.~Shakeri, Y.~Zhang, and N.~Hasani, ``Rail-only: A
  low-cost high-performance network for training llms with trillion
  parameters,'' 2024. [Online]. Available:
  \url{https://arxiv.org/abs/2307.12169}
\BIBentrySTDinterwordspacing

\bibitem{Gemini_Wang2023}
\BIBentryALTinterwordspacing
Z.~Wang, Z.~Jia, S.~Zheng, Z.~Zhang, X.~Fu, T.~S.~E. Ng, and Y.~Wang, ``Gemini:
  Fast failure recovery in distributed training with in-memory checkpoints,''
  in \emph{Proceedings of the 29th Symposium on Operating Systems
  Principles}.\hskip 1em plus 0.5em minus 0.4em\relax New York, NY, USA:
  Association for Computing Machinery, 2023, p. 364–381. [Online]. Available:
  \url{https://doi.org/10.1145/3600006.3613145}
\BIBentrySTDinterwordspacing

\bibitem{slurm}
A.~B. Yoo, M.~A. Jette, and M.~Grondona, ``Slurm: Simple linux utility for
  resource management,'' in \emph{Workshop on job scheduling strategies for
  parallel processing}.\hskip 1em plus 0.5em minus 0.4em\relax Springer, 2003,
  pp. 44--60.

\bibitem{Young_1974}
\BIBentryALTinterwordspacing
J.~W. Young, ``A first order approximation to the optimum checkpoint
  interval,'' \emph{Commun. ACM}, vol.~17, pp. 530--531, 1974. [Online].
  Available: \url{https://doi.org/10.1145/361147.361115}
\BIBentrySTDinterwordspacing

\bibitem{zhang-vldb2023}
T.~Zhang, K.~Liu, J.~Kosaian, J.~Yang, and R.~Vinayak, ``Efficient fault
  tolerance for recommendation model training via erasure coding,'' \emph{Proc.
  VLDB Endow.}, vol.~16, no.~11, p. 3137–3150, jul 2023.

\bibitem{gpu_age_aware_scheduling}
\BIBentryALTinterwordspacing
C.~Zimmer, D.~Maxwell, S.~McNally, S.~Atchley, and S.~S. Vazhkudai, ``Gpu
  age-aware scheduling to improve the reliability of leadership jobs on
  titan,'' in \emph{Proceedings of the International Conference for High
  Performance Computing, Networking, Storage, and Analysis}, ser. SC '18.\hskip
  1em plus 0.5em minus 0.4em\relax IEEE Press, 2019. [Online]. Available:
  \url{https://doi.org/10.1109/SC.2018.00010}
\BIBentrySTDinterwordspacing

\end{thebibliography}

\ifthenelse{\boolean{isarxivversion}}{{%
\arxivonly{
\newpage
\begin{appendix}

\subsection{Derivations of Expected ETTR\label{appendix:\ettrabbr{}}}
This section provides the derivation of expected ETTR for a cluster.
ETTR is defined for a single training run, but we can expect that there is variance even holding the run fixed due to the randomness of failures and scheduling.
Expected ETTR allows us to characterize typical job behavior if we were observing the job occurrence repeatedly.
We first define wallclock time $W = R + U + Q$ where $R$ is the productive training time of the job, $U$ is the unproductive scheduled time, and $Q$ is time the job spends unscheduled in the queue. 
Recall that $\text{\ettrabbr{}} = R / W$. We define $S = (U + Q) / R$ as an intermediate variable to help with deriving expectations for $\text{\ettrabbr{}} = 1 / (1 + S)$. By Jensen's inequality~\cite{HANSEN_2003}, $\mathbb{E}[\text{\ettrabbr{}}] \geq 1 / (1 + \mathbb{E}[S])$. If we hold productive runtime constant, we can take the expectation on both sides:

\begin{equation}
\mathbb{E}[S] = \frac{1}{R}\left(\mathbb{E}[U] + \mathbb{E}[Q]\right)
\end{equation}

The total queue time, $Q$, of a training run is the sum of the individual waiting or queueing times: $Q = q_0 + \sum_{j=1}^{N_{\text{int}}}q_j$ where $N_{\text{int}}$ is the number of job interruptions during the training run, $q_0$ is the initial wait time after submitting the job, and $q_j$ are the queue times after each interruption. If we assume wait times are independent and drawn i.i.d. from the same distribution $q_0,q_j \sim p(q)$, then $\mathbb{E}[Q] = (1 + \mathbb{E}[N_{\text{int}}])\mathbb{E}[q]$. We denote $\mathbb{E}[q]$ as $q$ hereafter.

Now we turn to unproductive training time, $U$, which is determined by job overheads and initialization costs, $u_j$, for the $j$-th interruption. The unproductive training time is $U = \sum_{j=0}^{N_{\text{int}}}\min(u_0 + N_{cp}w_{cp} + (W_j - t_{cp}), W_j)$, where $W_j$ is the wallclock time associated with the $j$-th job after the $j$-th interruption, $w_{cp}$ is the synchronous write cost of a checkpoint, $N_{cp}$ is the number of checkpoints written during the $j$-th job, and $t_{cp}$ is the time at which the final checkpoint write completed. We assume that upon each restart, we need to perform the same initialization tasks as during the first job instance ($u_0$), and we need to start training from the previous checkpoint.

If we assume that interruption time stamps are uncorrelated with checkpoint timestamps, and that checkpoint frequency is much higher than the interruption rate, $\mathbb{E}[W_j - t_{cp}] \approx \Delta t_{cp} / 2$ where $\Delta t_{cp}$ is the time interval between checkpoints. Note that if there are e.g. filesystem-related issues where correlations are expected between checkpoint writes and failures, $\mathbb{E}[W_j - t_{cp}]$ may approach $\Delta t_{cp}$.

We break down the number of job interruptions into two components: the number of preemptions $N_{\text{pre}}$, and the number of failures $N_f$. We ignore preemptions here to focus on high priority (non-preemptable) jobs. If we treat failures as occurring randomly on the node level at some rate $r_f$ and as being uncorrelated with each other, then the expected number of failures is $\mathbb{E}[N_f] = N_{\text{nodes}} r_f (R + \mathbb{E}[U])$.

If we consider the regime where $\Delta t_{cp}/2 + u_0 \ll (N_{\text{nodes}}r_f)^{-1}$, then $\mathbb{E}[U] = (\mathbb{E}[N_f] + 1)u_0 + \mathbb{E}[N_f]\Delta t / 2 + Rw_{cp}/\Delta t_{cp}$ and

\begin{equation}
\mathbb{E}[N_f] \approx RN_{\text{nodes}}r_f\left[\frac{\left(1 + \frac{u_0}{R} + \frac{w_p}{\Delta t_{cp}}\right)}{1 - N_{\text{nodes}}r_f\left(u_0 + \frac{\Delta t_{cp}}{2}\right)}\right]
\end{equation}

\begin{equation}
\mathbb{E}[S] \approx \frac{1}{R}\left((\mathbb{E}[N_f] + 1)\left(\mathbb{E}[q] + u_0\right) + \mathbb{E}[N_f] \frac{\Delta t_{cp}}{2} + \frac{Rw_{cp}}{\Delta t_{cp}}\right) 
\end{equation}

\begin{equation}
\mathbb{E}[\text{\ettrabbr{}}] \gtrsim (1 + \mathbb{E}[S])^{-1}
\end{equation}

Thus, the full expression for $\mathbb{E}[\text{\ettrabbr{}}]$ is
\begin{equation}
\mathbb{E}[\text{\ettrabbr{}}] \gtrsim \frac{1 - N_{\text{nodes}} r_f\left(u_0 + \frac{\Delta t_{cp}}{2}\right)}{1 + \frac{u_0 + q}{R} + \frac{w_{cp}}{\Delta t_{cp}} + N_{\text{nodes}} r_f q\left(1 + \frac{w_{cp}}{\Delta t_{cp}}  - \frac{\Delta t_{cp}}{2R}\right)}
\end{equation}

Optimal checkpointing intervals have been derived in many contexts, e.g. Daly-Young \cite{Daly_2006, Young_1974} and extensions \cite{Benoit_etal_2022, Bautista-Gomez_etal_2024,epoch_hardware_failures_2024}. We can find the checkpoint interval that maximizes \ettrabbr{} which involves solving for zeros of a cubic polynomial and in general depends on all parameters; however, a simpler approach is to make the same assumptions as Daly-Young ($\Delta t \gg u_0, w_{cp}, q$ and $R \gg u_0, q, \Delta t_{cp}, w_{cp}$) to obtain the canonical result:

\begin{equation}
    \Delta t_{cp}^* = \sqrt{\frac{2w_{cp}}{N_{\mathrm{nodes}}r_f}}
\end{equation}

Note that in practice there is often a maximum checkpointing frequency determined by e.g. step size time ($\mathcal{O}(10s)$ for SOTA LLM training), and that we have no ability to enforce utilization of any particular checkpointing interval.

For long-running, high priority jobs, where $u_0 + q + \Delta t_{cp} / 2 \ll R$, and in the presence of small queue time ($q\approx 0$)
\begin{equation}
\mathbb{E}[\text{\ettrabbr{}}] \approx \frac{1 - N_{\text{nodes}}r_f\left(u_0 + \frac{\Delta t_{cp}}{2}\right)}{1 + \frac{w_{cp}}{\Delta t_{cp}}}
\end{equation}

or 

\begin{equation}
\mathbb{E}[\text{\ettrabbr{}}] \approx \frac{1 - N_{\text{nodes}}r_f\left(u_0 + \sqrt{\frac{w_{cp}}{2N_{\mathrm{nodes}}r_f}}\right)}{1 + \sqrt{\frac{N_{\text{nodes}}r_fw_{cp}}{2}}}
\end{equation}

if using Daly-Young checkpoint intervals.

\end{appendix}
}
}}{{}}

\end{document}